\documentclass[referee]{aa}

\usepackage[dvips]{graphicx}
\usepackage{epsfig}
\usepackage{txfonts}
\usepackage{natbib}

\begin{document}

\title{Magnetic reconnection at the termination shock in a striped
  pulsar wind.}

\author{J\'er\^ome P\'etri \inst{1} \and Yuri Lyubarsky \inst{2} }

\offprints{J. P\'etri}

\institute {Max-Planck-Institut f\"ur Kernphysik, Saupfercheckweg 1,
  69117 Heidelberg, Germany \and Department of Physics, Ben-Gurion
  University, P.O.Box 653, Beer-Sheva 84105, Israel.}

\date{Received / Accepted}

\titlerunning{Magnetic reconnection in a striped pulsar wind}

\authorrunning{P\'etri \and Lyubarsky}

\abstract
%
{Most of the rotational luminosity of a pulsar is carried away by a
  relativistic magnetised wind in which the matter energy flux is
  negligible compared to the Poynting flux. However, observations of
  the Crab nebula for instance clearly indicate that most of the
  Poynting flux is eventually converted into ultra-relativistic
  particles. The mechanism responsible for transformation of the
  electro-magnetic energy into the particle energy remains poorly
  understood. Near the equatorial plane of an obliquely rotating
  pulsar magnetosphere, the magnetic field reverses polarity with the
  pulsar period, forming a wind with oppositely directed field lines.
  This structure is called a striped wind; dissipation of alternating
  fields in the striped wind is the object of our study.}
{The aim of this paper is to study the conditions required for
  magnetic energy release at the termination shock of the striped
  pulsar wind.  Magnetic reconnection is considered via analytical
  methods and 1D relativistic PIC simulations. }
{An analytical condition on the upstream parameters for partial and
  full magnetic reconnection is derived from the conservation laws of
  energy, momentum and particle number density across the relativistic
  shock. Furthermore, by using a 1D relativistic PIC code, we study in
  detail the reconnection process at the termination shock for
  different upstream Lorentz factors and magnetisations.}
{We found a very simple criterion for dissipation of alternating
  fields at the termination shock, depending on the upstream
  parameters of the flow, namely, the magnetisation~$\sigma$, the
  Larmor radius $r_{\rm B}$ and the wavelength~$l$ of the striped
  wind. The model depends also on a free parameter~$\xi>1$, which is
  the ratio of the current sheet width to the particle Larmor radius.
  It is found that for $\sigma \gg l / \xi \, r_{\rm B}$, all the
  Poynting flux is converted into particle energy whereas for $\sigma
  \ll (l / \xi \, r_{\rm B} )^{2/3}$, no dissipation occurs. In the
  latter case, the shock can be accurately described by the ideal MHD
  shock conditions.  Finally, 1D relativistic PIC simulations confirm
  this prediction and enable us to fix the free parameter~$\xi$ in the
  analytical model.}
{ Alternating magnetic fields annihilate easily at relativistic highly
  magnetised shocks. In plerions, our condition for full magnetic
  dissipation is satisfied at the termination shock so that the
  Poynting flux may be converted into ultra-relativistic particles not
  in the pulsar wind but just at the termination shock.  The
  constraints are more severe for the intra-binary shocks in double
  pulsar systems.  Available models explaining observations require
  low magnetisation in the downstream flow.  The condition that the
  magnetic field dissipates at the intra-binary shock implies an upper
  limit on the pair multiplicity in the pulsar wind~$\kappa$.  We
  found $\kappa \lesssim {\rm few} \times 10^4$ for PSR~1259-63 and
  PSR~1957+20. In the double pulsar PSR~0737-3039, the radio emission
  from the pulsar B is modulated with the period of the pulsar A,
  which implies that the striped structure is not erased completely;
  this gives a lower limit for $\kappa \gtrsim 310$.}


\maketitle

\section{INTRODUCTION}

Relativistic shock fronts and currents sheets in relativistic flows
play an important role in astrophysical models of gamma-ray bursts,
for a review see \cite{2005RvMP...76.1143P}, jets in active galactic
nuclei and pulsar winds, see for instance \cite{2005RMxAC..23...27M}
and \cite{2005PPCF...47B.719K}. The underlying plasma is probably
composed of electrons, positrons and/or protons, whose temperature may
be relativistic, i.e. comparable or larger than their rest mass
energy. A shock front is created whenever a fast flow encounters the
interstellar or intergalactic medium.  Relativistic effects become
important when the post shock temperature is so high that the speed of
particles approaches the speed of light~$c$ or when the bulk velocity
of the flow is close to~$c$. Shock acceleration is used to explain the
observed radiation from gamma-ray bursts or active galactic nuclei.
In this paper, we focus on relativistic shocks arising form the
interaction of the pulsar wind with its surrounding medium.  For a
detailed review on theoretical aspects on pulsar wind and plerions,
see \cite{2005xrrc.procE5.02L}.  We briefly recall the main issue in
this introduction.

It is widely assumed that most of the rotational energy of a pulsar is
carried away in the form of an ultra-relativistic magnetised wind. The
outflow is dominated by the magnetic field in the sense that the
energy carried away by the plasma remains small compared to the
Poynting flux. This is usually described by the magnetization
parameter~$\sigma$, the ratio of the Poynting flux to the particle
energy flux, which is very large, $\sigma \gg 1$.  Therefore the total
power lost by the pulsar may be conveniently estimated by the loss
rate of the rotating magnetic dipole in vacuum,
\citep{1991tnsm.book.....M}.  In the general case of an oblique
rotator, energy loss can be thought as being shared between the steady
axisymmetric component and one oscillating with the period of the
pulsar, the ratio being determined by the angle between the rotational
and the magnetic axes.  \cite{1971CoASP...3...80M} pointed out that
such waves, which have a phase speed less than that of light, should
evolve into regions of cold magnetically dominated plasma separated by
narrow hot current sheets.  This structure is called a striped pulsar
wind and has originally been introduced by \cite{1990ApJ...349..538C}
and \cite{1994ApJ...431..397M}.

\begin{figure}[htbp]
  \centering \includegraphics[scale=0.6]{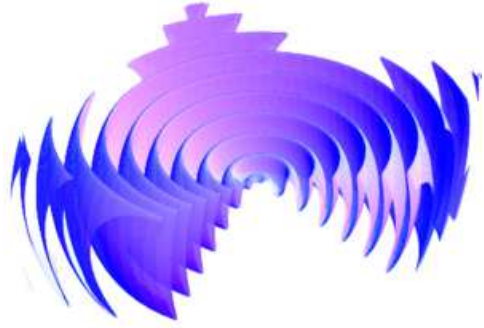}
  \caption{Structure of the current sheet in the oblique rotator.
    The rotating neutron star is located at the centre. When crossing
    the current sheet, the magnetic polarity is reversed.  Moreover,
    this structure is expanding radially outwards with a constant
    speed close to the light velocity.}
  \label{fig:StripedWind}
\end{figure}

The structure of this striped wind can be explained as follows. In the
aligned rotator, a radially outflowing stream of particles opens up
the dipolar magnetic field lines crossing the light cylinder.
Asymptotically, the stream lines become radial and can be described by
the so-called split monopole solution \citep{1973ApJ...180L.133M}. It
consists of two half magnetic monopoles with equal magnitude but
opposite sign separated by the equatorial plane.  When crossing this
equatorial plane, the polarity of the magnetic field reverses,
implying a surface current density, i.e. the current sheet. If the
rotation axis is tilted with respect to the magnetic moment, the
current sheet oscillates around the equatorial plane and develops a
wave structure, expanding radially at a speed close to the light
velocity. In a radial outflow, the amplitude of the oscillations grows
linearly with radius and at large distances can be approximated
locally by spherical current sheets separating the stripes of
magnetised plasma with opposite magnetic polarity. The exact solution
for the oblique split monopole in the ideal MHD approximation has been
found by \cite{1999A&A...349.1017B}.  Simulations by
\cite{2006ApJ...648L..51S} in the force-free limit show that beyond
the light cylinder, the wind from an oblique dipole pulsar
magnetosphere is similar to that from the split monopole.  The
structure of the current sheet in the striped wind is shown in
Fig.~\ref{fig:StripedWind}.  Note that the oscillating current sheet
is found in the solar wind so that this structure is generic.

While expanding into the nebula, the pulsar wind terminates at a
standing shock located at a distance where the confining pressure of
the nebula balances the ram pressure of the wind, this is called the
termination shock, \citep{1974MNRAS.167....1R, 1984ApJ...283..694K,
  1984ApJ...283..710K, 1987ApJ...321..334E}.  At the shock
front, the energy in the wind is released into ultra-relativistic
particles responsible for the observed radiation.  Observations of the
interaction between the wind and the neighboring nebula clearly
indicate that the electromagnetic energy of the wind is largely
converted into particle kinetic energy.  However, the conversion
mechanism and the corresponding acceleration process still remain
unclear.

Indeed, the analysis done by different authors,
\citep{1994PASJ...46..123T, 1998MNRAS.299..341B, 1998ApJ...505..835C,
  1999MNRAS.305..211B, 2001ApJ...562..494L} shows that in an
ultra-relativistic, radial wind, the electric force compensates the
magnetic tension so that the flow could not be accelerated
significantly.  This means that the electromagnetic energy is not
transferred to the plasma. To sum up, in the region where the wind is
launched, the magnetisation should be high, $\sigma\gg1$, whereas just
beyond the termination shock, it should be small, $\sigma\ll1$.  But
according to the previous results on MHD wind collimation and
acceleration, a significant decrease in $\sigma$ is forbidden.
Therefore, there is a contradiction which is known as the
$\sigma$-problem.  A promising alternative solution to convert the
Poynting flux to particle kinetic energy is investigated in this
paper. We demonstrate that, under certain conditions, magnetic
reconnection in the striped wind occurs when it crosses the
termination shock.

Both observations of the inner region of the Crab nebula done by
\cite{2000ApJ...536L..81W} and the solution by
\cite{1999A&A...349.1017B} show that most of the energy carried by the
wind is transported in the equatorial plane of the pulsar wind. In
this region, energy is transferred predominantly by the alternating
magnetic field. So dissipation of alternating fields in the striped
wind could be the main energy conversion mechanism in pulsars.  It was
recognised by \cite{1975Ap&SS..32..375U} and
\cite{1982RvMP...54....1M} that the amplitude of the magnetic
oscillations decreases with the distance as $r^{-1}$ whereas the
particle number density sustaining these waves falls off as $r^{-2}$
where $r$ is the radius in spherical coordinates. At a given point,
the charge carriers become insufficient to maintain the required
current and the alternating field annihilates.
\cite{1990ApJ...349..538C} considered magnetic reconnection in the
striped wind and was lead to the same conclusion. However, the flow is
accelerated during the process of magnetic dissipation, which dilates
the timescale of the wave decay such that the magnetisation remains
high at the termination shock, \citep{2001ApJ...547..437L}.
Therefore, the wind enters the termination shock still dominated by
Poynting flux unless the annihilation rate is nearly to the causal
limit, \citep{2003ApJ...591..366K}. However, when the flow enters the
shock, the plasma is compressed, which could result in the forced
annihilation of the alternating magnetic fields.  In this paper, we
address annihilation of alternating fields at a relativistic shock in
electron-positron plasma.

In \cite{2003MNRAS.345..153L}, the reconnection of the magnetic field
at the termination shock was studied phenomenologically by introducing
a fraction $\eta$ of the magnetic energy dissipated at the shock.
Relativistic MHD shock fronts with dissipation have already been
studied by \cite{1997ApJ...488...69L}. The jump conditions for a
relativistic perpendicular ideal MHD shock with arbitrary
magnetisation has been investigated by \cite{2005ApJ...628..315Z} in
the context of GRBs.  In the present paper, we use a kinetic
description of the plasma in the striped wind.  First we find jump
conditions assuming that the thickness of the current sheets
downstream of the shock is scaled as the particle Larmor radius; this
gives us the fraction of the dissipated magnetic energy as a function
of upstream parameters and of a phenomenological parameter, $\xi>1$,
which is defined as the ratio of the sheet width to the Larmor radius.
Then we perform particle-in-cell (PIC) simulations of the shock in the
striped wind (such simulations were shortly described in
\cite{2005AdSpR..35.1112L}) and find the parameter $\xi$ from the
simulation results.  We show that the alternating fields are easily
annihilated at the shock front so that the electromagnetic energy of
the pulsar wind could be readily converted into the plasma energy at
the termination shock. This could provide a solution to the
$\sigma$-problem.

The paper is organised as follows. In Sect.~\ref{sec:Condition},
we present the full system for the jump conditions of the averaged
quantities in the striped wind.  In Sect.\ref{sec:Solutions}, we
solve the system analytically. We derive an analytical condition
for magnetic annihilation for an ultra-relativistic and strongly
magnetised flow.  The results are then checked and extended by
numerically solving the jump conditions at the termination shock
for the average parameters. In Sect.\ref{sec:Pic}, we perform
several 1D PIC simulations of the striped wind by varying the
parameters in the upstream plasma like the Lorentz factor and the
magnetisation.  Two typical situations are presented, the first
one demonstrating that no magnetic energy is released at all and
the second one showing full dissipation of the Poynting flux which
is converted into particle heating.  An empirical law for
dissipation is then derived from the full set of PIC simulations.
The results are compared to those obtained in
Sect.~\ref{sec:Solutions}. In Sect.\ref{sec:Wind}, the results are
applied to pulsars in binary systems. On one hand, magnetic
dissipation at the termination shock implies an upper limit on the
pair multiplicity factor~$\kappa$. Applications are presented for
two binary pulsars, PSR~1259-63 and PSR~1957+20. On the other
hand, an absence of significant dissipation imposes a lower limit
on  $\kappa$. This is applied to the double pulsar PSR~0737-3039.
The conclusion are presented in Sect.~\ref{sec:Conclusion}.

\section{JUMP CONDITIONS IN THE SHOCK}
\label{sec:Condition}

In this section, we derive the criterion for dissipation of
alternating field at the shock by considering the jump conditions for
the averaged quantities in the flow, namely, the conservation of the
energy, momentum and the particle number density.  We assume that the
striped structure survives in the downstream flow and find the
condition of the total dissipation of the alternating fields from the
condition that the thickness of the current sheet downstream of the
shock approaches the thickness of the stripes.

\subsection{Description of the wind flow}

Let us consider a one dimensional striped pulsar wind entering the
termination shock. By convention, we refer to quantities in the shock
frame with unprimed letters, while in the proper frame of each plasma,
quantities are denoted by a prime. However, thermodynamical quantities
such as pressure~$p$, temperature~$T$, internal energy~$e$ and
enthalpy~$w$, are always expressed in the proper frame, so dropping
primes should not lead to any confusion for these quantities.

The striped wind propagates in the $x$-direction with relativistic
velocity and possesses an alternating magnetic field directed
along the $z$-axis. Moreover, in this and the next sections we
assume that its average over one period of the striped wind
vanishes. Quantities upstream, i.e. before crossing the shock
discontinuity are subscripted by~1 whereas quantities downstream
are subscripted by~2.

Let us describe the state of the incoming plasma. In the shock frame,
the upstream Lorentz factor of the ultra-relativistic wind is
$\Gamma_1 \gg 1$. The current sheets are made of a hot, unmagnetised
plasma with density~$n_{\rm h1}$ and a temperature~$T_{\rm h1}$.  The
distance separating the middle of two successive current sheets is
denoted by~$l_1$ (half a wavelength of the wind) and their thickness
$\Delta_1$ is much less than half a wavelength,~$\Delta_1\ll l_1$.
The magnetised part of the wind is cold, $T_{\rm c1}=0$, has a
density~$n_{\rm c1}$ and a magnetic field strength~$B_1$. Moreover, the
wind is strongly magnetised such that the magnetisation parameter
upstream defined by
\begin{equation}
  \label{eq:Magnetisation1}
  \sigma_1 = \frac{B_1^2}{\mu_0 \, \Gamma_1^2 \, w_{\rm c1}} =
  \frac{B_1^2}{\mu_0 \, \Gamma_1 \, n_{\rm c1} \, m \, c^2}
\end{equation}
is very high, $\sigma_1 \gg 1$. The enthalpy is simply given by the
rest mass energy of the cold ultra-relativistic particles, $w_{\rm
  c1}=n_{\rm c1}'\,m\,c^2$. The speed of light is~$c$ and~$m$ is the
mass of the leptons composing the wind, actually electrons and
positrons.  We neglect the enthalpy contribution from the hot current
sheets because their thickness is assumed to be very small.

Downstream, the wind is decelerated to a Lorentz factor~$\Gamma_2$ and
compressed such that the distance separating two successive current
sheets shrinks to a length~$l_2$.  When compressed, the cold
magnetised component of the wind is heated to a temperature~$T$ given
in the limit of high magnetisation, $\sigma_1 \gg 1$, and high Lorentz
factor, $\Gamma_1 \gg 1$, by \citep{1984ApJ...283..694K} (see also
appendix~\ref{sec:AnnexeA})~:
\begin{equation}
  \label{eq:Taval}
  \frac{k_B \, T}{m \, c^2} = \frac{1}{8} \, \frac{\Gamma_1}{\Gamma_2}
\end{equation}
$k_B$ is the Boltzmann constant.  We assume here that the shock width
is much less than the wavelength of the wind. When the shock is
between the sheets, ideal MHD shock applies and Eq.~(\ref{eq:Taval})
is locally valid. Note also that the shock velocity in the cold part
is close to the speed of light~$c$ and different from the shock
velocity in the hot part. Quantities are evaluated in the frame where
the shock is at rest on average.  Eq.~(\ref{eq:Taval}) is valid in any
frame.

Note that even if the upstream flow is a pure entropy wave with a
constant magnetic field between the sheets, fast magnetosonic waves
should be generated beyond the shock.  Therefore the structure
downstream is not steady in the proper frame; there should be
oscillations. We consider quantities averaged over the wave period and
neglect contribution of these oscillations into the fluxes.

Pressure balance between gaseous and magnetic part does therefore
apply on both sides of the discontinuity.  The gaseous pressures are
given by~:
\begin{eqnarray}
  \label{eq:P1}
  p_1 & = & n_{\rm h1}' \, k_B \, T_{\rm h1} \\
  \label{eq:P2}
  p_2 & = & n_{\rm h2}' \, k_B \, T_{\rm h2} \\
  \label{eq:Pc2}
  p_{\rm c2} & = & n_{\rm c2}' \, k_B \, T =
  \frac{1}{8} \, \frac{\Gamma_1}{\Gamma_2} \, n_{\rm c2}' \, m \, c^2
\end{eqnarray}
Note that the ``cold'' part of the wind is heated to relativistic
temperature downstream and therefore also contributes to the gaseous
pressure via the term~$p_{\rm c2}$, where the temperature is given
according to Eq.~(\ref{eq:Taval}). In other words, upstream we have
\begin{equation}
  \label{eq:EqPressAmont}
  p_1= \frac{B_1^2}{2\,\mu_0\,\Gamma_1^2}
\end{equation}
whereas downstream, taking into account the heated cold component, we
find
\begin{equation}
  \label{eq:EqPressAval}
  p_2=
  p_{c2}+ \frac{B_2^2}{2\,\mu_0\,\Gamma_2^2}
\end{equation}
Due to the Lorentz length contraction, the densities in the proper
frame and in the shock frame are related by~$n = \Gamma \, n'$. On the
other hand, the Lorentz transformation of the magnetic field is $B =
\Gamma \, B'$.

\subsection{Jump conditions for average quantities}

We now write down the MHD jump conditions for the quantities averaged
over one wavelength of the wind.  Following the procedure described in
\cite{2003MNRAS.345..153L} for a perpendicular MHD shock, the
conservation laws can be cast in a form similar to those for a
relativistic hydrodynamical flow. Noting that in the cold magnetised
part, the field is frozen into the plasma
\begin{equation}
  \label{eq:ConservFluxMag}
  \frac{B_1}{\Gamma_1 \, n'_{c1}} = \frac{B_2}{\Gamma_2 \, n'_{c2}} \equiv
  \frac{B_1}{n_{c1}} = \frac{B_{c2}}{n_{c2}}
\end{equation}
such that the magnetic field becomes a function of the density, one
can introduce the effective pressure and enthalpy by~:
\begin{eqnarray}
  \label{eq:Peff}
  \mathcal{P}_i & = & p_i + \frac{B_i^2}{2 \, \mu_0 \, \Gamma_i^2} \\
  \label{eq:Weff}
  \mathcal{W}_i & = & w_i + \frac{B_i^2}{     \mu_0 \, \Gamma_i^2}
\end{eqnarray}
Now the jump conditions at the shock discontinuity are given by
\begin{itemize}
\item conservation of particle number density~:
  \begin{equation}
    \label{eq:ConservPart}
    \langle \Gamma_1 \, \beta_1 \, n_1' \rangle =
    \langle \Gamma_2 \, \beta_2 \, n_2' \rangle
    \equiv \langle \beta_1  \, n_1\rangle  =
    \langle \beta_2 \, n_2\rangle ;
  \end{equation}
\item conservation of total energy~:
  \begin{equation}
    \label{eq:ConservEnerg}
    \langle\Gamma_1^2 \, \beta_1 \, \mathcal{W}_1\rangle =
    \langle\Gamma_2^2 \, \beta_2 \, \mathcal{W}_2\rangle;
  \end{equation}
\item conservation of total momentum~:
  \begin{equation}
    \label{eq:ConservImpuls}
    \langle\Gamma_1^2 \, \beta_1^2 \, \mathcal{W}_1 + \mathcal{P}_1\rangle =
    \langle\Gamma_2^2 \, \beta_2^2 \, \mathcal{W}_2 +
    \mathcal{P}_2\rangle.
  \end{equation}
\end{itemize}
Here $\beta_{1,2}c$ are the upstream and downstream velocities,
correspondingly and the angular brackets, $\langle \rangle$, mean
averaging over the wave period, which means spatial averages.  We
assume that the downstream flow is settled into an equilibrium pattern
moving with constant velocity, i.e. that only an entropy wave is
presented downstream. Then time averages are expressed via spatial
averages; e.g. the average particle flux is
\begin{equation}
  \label{eq:aver}
  \frac {1}{T} \, \int_0^T n \, \beta \, dt = \frac {1}{T} \, \int_0^l
  n \, dx = \frac {\beta}{l} \, \int_0^l n \, dx.
\end{equation}
As the time-averaged particle flux (the left-hand side of
Eq.~(\ref{eq:aver})) is constant and the total number of particles
within the wavelength ($ \int_0^l n \, dx$) is conserved,
Eq.~(\ref{eq:aver}) yields a useful relation between the upstream and
downstream wavelengths:
\begin{equation}
  \label{eq:wavelength}
  \frac{\beta_1}{l_1}=\frac{\beta_2}{l_2}.
\end{equation}

Of course, our assumption is not exactly correct; when an entropy wave
impinges on the shock, both entropy and fast magnetosonic waves
generally appear in the downstream flow.  One could assume that the
shock width is small as compared to the sheet width; then one could
apply Rankine-Hugoniot relations locally and find the time-dependent
structure of the downstream flow.  However, such an assumption is too
stringent; it could be valid in fact only when dissipation is
negligible. In this paper, we look for a criterion for complete
dissipation of the alternating magnetic field. This criterion is found
by extrapolating our analytical model to the case when the downstream
Larmor radius becomes comparable to the wave period so that the
structure assumed in the model is already destroyed.  Therefore this
criterion is anyway very rough.  We believe that such a rough
criterion should be independent of the fine structure of the
downstream flow, like fast magnetosonic oscillations.  PIC simulations
indeed show that our criterion for dissipation is roughly correct so
that our model is viable. Detailed study of the shock interaction with
the current sheets in case of partial dissipation, as well as
generation of fast magnetosonic waves in the downstream flow, is the
subject of our future work.

In the hot phases of the plasma, we use the ultra-relativistic
equation of state such that $w = 4 \, p$.  Therefore, the average
effective enthalpies upstream and downstream are
\begin{eqnarray}
  \label{eq:W1}
  \langle \mathcal{W}_1 \rangle & = & 4 \, p_1 \, \delta_1 + ( 1 - \delta_1 ) \,
  \left[ n_{\rm c1}' \, m \, c^2 + \frac{B_1^2}{\mu_0\,\Gamma_1^2} \right] \\
  \label{eq:W2}
  \langle \mathcal{W}_2 \rangle & = & 4 \, p_2 \, \delta_2 + ( 1 - \delta_2 ) \,
  \left[ 4 \, p_{\rm c2} + \frac{B_2^2}{\mu_0\,\Gamma_2^2} \right]
\end{eqnarray}
For the remaining of this paper, we introduce the relative thickness
of a current sheet by $\delta_i = \Delta_i / l_i$, ($i=1,2$). Making
use of the pressure balance condition Eq.(\ref{eq:EqPressAmont}), one
can express the upstream enthalpy via the magnetisation parameter
Eq.~(\ref{eq:Magnetisation1}) as
\begin{equation}
  \langle \mathcal{W}_1 \rangle
  =n'_{\rm c1} \, m \, c^2 \,
  [ 1 - \delta_1 + ( 1 + \delta_1 ) \, \sigma_1 ].
\end{equation}
The downstream pressure balance Eq.(\ref{eq:EqPressAval}) could be
also expressed via $\sigma_1$ taking into account that between the
current sheets, the magnetic field is frozen into the plasma,
Eq.~(\ref{eq:ConservFluxMag}):
  \begin{equation}
  \label{eq:Balance}
  p_2 = \frac 18 \, \frac{\Gamma_1}{\Gamma_2} \, n'_{c2} \,
  m \, c^2 \, \left( 1 + 4 \, \frac{n_{c2}}{n_{c1}} \, \sigma_1 \, \right).
  \end{equation}
Then the downstream enthalpy is written as
\begin{equation}
  \label{eq:W2s}
  \langle \mathcal{W}_2 \rangle = \frac{1}{2} \,
  \frac{\Gamma_1}{\Gamma_2} \, n'_{\rm c2} \, m \, c^2
  \left[
  1 + ( 1 + \delta_2 ) \, 2 \, \frac{n_{\rm c2}}{n_{\rm c1}} \, \sigma_1
  \right].
\end{equation}
Now the conservation of particle number density, energy and momentum
simplifies respectively into~:
\begin{eqnarray}
  \label{eq:CsvPart}
  \beta_1 \, \left[ ( 1 - \delta_1 ) \, n_{\rm c1} + \delta_1 \, n_{\rm h1} \right] & = &
  \beta_2 \, \left[ ( 1 - \delta_2 ) \, n_{\rm c2} + \delta_2 \, n_{\rm h2} \right] \\
  \label{eq:CsvEnerg}
  2 \, \beta_1 \, n_{\rm c1} \, \left[ 1 - \delta_1 +
    ( 1 + \delta_1 ) \, \sigma_1 \right] & = &
  \beta_2 \, n_{\rm c2} \, \left[
    1 + ( 1 + \delta_2 ) \, 2 \, \frac{n_{\rm c2}}{n_{\rm c1}} \, \sigma_1
  \right] \\
  \label{eq:CsvImpuls}
  n_{\rm c1} \, \left[ \Gamma_1 \, \beta_1^2 \, \left[ 1 - \delta_1 +
      ( 1 + \delta_1 ) \, \sigma_1 \right] + \frac{\sigma_1}{2\,\Gamma_1} \right] & = &
  n_{\rm c2} \, \left\{ \Gamma_2 \, \beta_2^2 \,  \left[
      1 + ( 1 + \delta_2 ) \, 2 \, \frac{n_{\rm c2}}{n_{\rm c1}} \, \sigma_1
    \right] + \frac{1}{4\,\Gamma_2} +
    \frac{n_{\rm c2} \, \sigma_1}{n_{\rm c1} \, \Gamma_2} \right\}
  \, \frac{1}{2} \, \frac{\Gamma_1}{\Gamma_2} \nonumber \\
 & &
\end{eqnarray}
The striped wind downstream is described by four parameters, namely,
the particle number density in the hot (unmagnetised) and cold
(magnetised) part, $n_{\rm h2}$ and $n_{\rm c2}$, the Lorentz factor
$\Gamma_2$ and the relative current sheet thickness~$\delta_2$. The
available three equations (\ref{eq:CsvPart}), (\ref{eq:CsvEnerg}),
(\ref{eq:CsvImpuls}) should be complemented by one more equation; some
assumption about microphysics of the reconnection process is necessary
to close the system.

The most natural assumption is that in the course of the reconnection
process the current sheet thickness scales as the Larmor radius of the
particle in the sheet; then the system is closed by introducing a
parameter~$\xi>1$ such that the downstream current sheet thickness in
the proper frame is defined by
\begin{equation}
  \label{eq:Delta2p}
  \Delta_2' = \xi \, r_{B}'
\end{equation}
The Larmor radius in the downstream plasma frame is given by~:
\begin{equation}
  \label{eq:RLarmor}
  r_{B}' \approx \frac{k_B \, T_{\rm h2}}{|q| \, B_2' \, c} =
  \frac{\Gamma_2 \, k_B \, T_{\rm h2}}{|q| \, B_2 \, c}
\end{equation}
$q$ is the charge of a lepton (electron or positron), $q=\pm e$.
Expressing the temperature via the pressure Eq.~(\ref{eq:Pc2}) and
substituting the pressure balance condition Eq.~(\ref{eq:Balance}),
one can write Eq.~(\ref{eq:Delta2p}) as
\begin{equation}
  \label{eq:Del2}
  \Delta'_2 = \frac{\xi}8 \left(1+4\frac{n_{c2}}{n_{c1}}\sigma_1\right) \,
  \frac{n_{\rm c1}}{n_{\rm h2}}
  \, r_{0};
\end{equation}
where we introduced the Larmor radius related to the upstream bulk
velocity~$\beta_1\,c$ and defined by
\begin{equation}
  \label{eq:Larmoream}
  r_{\rm 0} = \frac{\Gamma_1 \, m \, c}{|q| \, B_1}.
\end{equation}
This quantity could be deduced from the upstream parameters as
\begin{equation}
  \label{eq:Larmor1}
  r_{0} = \frac{1}{|q|} \, \sqrt{\frac{\Gamma_1 \, m}{\mu_0 \, \sigma_1 \, n_{\rm
  c1}}}.
\end{equation}
Note that the expression~(\ref{eq:Larmoream}) is not the true Larmor
radius of the particles in the upstream flow, the last being dependent
on the plasma temperature. Transforming to the shock frame,
$\Delta_2=\Delta'_2/\Gamma_2$, dividing by $l_2$ and making use of
Eq.~(\ref{eq:wavelength}), one can finally write the closure condition
as
\begin{equation}
  \label{eq:del2}
  \delta_2 = \frac{\xi}{8} \, \left( 1 +
    4 \, \frac{n_{c2}}{n_{c1}} \, \sigma_1 \right) \,
  \frac{n_{\rm c1} \, \beta_1}{n_{\rm h2} \, \beta_2}
  \, \frac{r_{0}}{\Gamma_2 \, l_1}.
\end{equation}

\section{SOLUTIONS OF THE JUMP CONDITIONS}
\label{sec:Solutions}

The set of equations
(\ref{eq:CsvPart})-(\ref{eq:CsvEnerg})-(\ref{eq:CsvImpuls})-(\ref{eq:del2})
is first solved analytically under the assumption that the fraction of
the dissipated magnetic energy is small so that the downstream current
sheet remains narrow and the downstream flow remains
ultra-relativistic and strongly magnetised. Extrapolating of the
obtained asymptotics to the case when the thickness of the current
sheet in the downstream flow becomes comparable to the wavelength, we
obtain an analytical criterion for the full dissipation of the
alternating magnetic fields at the shock front. More general
conditions are then investigated by solving the system numerically.

\subsection{Analytical asymptotic solution}

We are interested in a highly relativistic, Poynting dominated
upstream flow so that $\sigma_1 \gg 1$, $\Gamma_1 \gg 1$, $\delta_1\ll
1$. In the ideal MHD, such a flow would remain highly relativistic
downstream of the shock, $\Gamma_2=\sqrt{\sigma_1}$,
\cite{1984ApJ...283..694K} and appendix~\ref{sec:AnnexeA}. When some
fraction of the magnetic energy is dissipated at the shock front, the
velocity of the downstream flow decreases and reaches $\beta_2=1/3$
when the magnetic field is completely dissipated
\citep{2003MNRAS.345..153L}. In this subsection, we assume that only a
small fraction of the magnetic energy is dissipated so that downstream
of the shock, the current sheets remain narrow, $\delta_{2} \ll 1$ and
the flow remains ultra-relativistic, $\Gamma_{2} \gg 1$. We solve
Eqs.~(\ref{eq:CsvPart})-(\ref{eq:CsvEnerg})-(\ref{eq:CsvImpuls})-(\ref{eq:del2})
assuming that $1/\sigma_1$, 1/$\Gamma_{1,2}$ and $\delta_{1,2}$ are
small. In the zeroth order approximation in these parameters (when all
of them are equal to zero), the three equations
(\ref{eq:CsvPart})-(\ref{eq:CsvEnerg})-(\ref{eq:CsvImpuls}) are
reduced to the same equality
\begin{equation}
  \label{eq:Zero}
  n_{\rm c1} = n_{\rm c2}
\end{equation}
so the system is nearly degenerate. It is therefore necessary to
retain higher order terms. For a while we do not assume any relations
between these parameters and just expand
Eqs.~(\ref{eq:CsvPart})-(\ref{eq:CsvEnerg})-(\ref{eq:CsvImpuls}) to
the first non-vanishing order in each of them independently. In the
obtained equations, both the zero and the first order terms are
presented.  Dealing with equations containing terms of different order
is difficult. One can simplify the problem if one uses
Eq.~(\ref{eq:CsvEnerg}) in order to eliminate the zeroth order terms
from Eq.~(\ref{eq:CsvPart}) and Eq.~(\ref{eq:CsvImpuls}),
correspondingly; then one gets two equations containing only small
order terms; these equations could be complemented by the zeroth order
equation (\ref{eq:Zero}) and also by Eq.~(\ref{eq:del2}) in the zeroth
order.

Expanding Eq.~(\ref{eq:CsvPart}) to first non-vanishing order in small
parameters we find
\begin{equation}
  \label{eq:nc1-nc2}
  n_{\rm c1} - n_{\rm c2} = \delta_1 \, ( n_{\rm c1} - n_{\rm h1} ) -
  \delta_2 \, ( n_{\rm c2} - n_{\rm h2} ) + \frac{n_{\rm c1}}{2 \, \Gamma_1^2} -
  \frac{n_{\rm c2}}{2 \, \Gamma_2^2}
\end{equation}
We put the zeroth order terms to the left hand side and the small
terms to the right hand side; one can see that the difference $n_{\rm
  c1} - n_{\rm c2}$ is small. Now let us divide
Eq.~(\ref{eq:CsvEnerg}) by $\sigma_1$ so that the leading order terms
be of zeroth order; then expanding in the small parameters to the
first non-vanishing order yields
\begin{equation}
\label{eq:gam}
  2 \, n_{\rm c1} \, \left[ 1 + \delta_1 +\frac 1{\sigma_1}  -
  \frac{n_{\rm c1}}{2\Gamma_1^2}\right] =
  n_{\rm c2} \, \left[ 2(1 + \delta_2 ) \,
    \frac{n_{\rm c2}}{n_{\rm c1}}+\frac 1{\sigma_1}  -
    \frac{n_{\rm c2}^2}{n_{\rm c1}} \, \frac{1}{\Gamma_2^2} \right]. \\
\end{equation}
One sees that neglecting small order terms would result again to
Eq.~(\ref{eq:Zero}) therefore the small terms should be retained.
Substituting Eq.~(\ref{eq:nc1-nc2}) into the zeroth order terms and
Eq.~(\ref{eq:Zero}) into the rest of the terms, one gets
\begin{equation}
  \label{eq:frac}
  \frac{1}{2\,\sigma_1} + \frac{1}{2\,\Gamma_1^2} - \frac{1}{2\,\Gamma_2^2} =
  \delta_2 \, ( 3 - \frac{2}{Z_2} ) - \delta_1 \, ( 3 - \frac{2}{Z_1} )
\end{equation}
Here, we introduced the fraction of cold to hot particle densities by
\begin{equation}
  \label{eq:Z}
  Z = \frac{n_c}{n_h}
\end{equation}

In order to eliminate the zeroth order terms from
Eqs.~(\ref{eq:CsvEnerg}) and (\ref{eq:CsvImpuls}), one can simply
extract one of them from another because the zeroth order terms in
these equations are the same (those containing $\Gamma_2 \,
\sigma_1$). Making use of Eq.~(\ref{eq:Zero}) in the rest of the
terms, we arrive at
\begin{equation}
  \label{eq:Gamma1}
  \delta_1 + \frac{1}{\sigma_1} - \frac{1}{4\,\Gamma_1^2} =
  \frac{\Gamma_1^2}{\Gamma_2^2} \, \left[ \delta_2 +
    \frac{1}{4 \, \sigma_1} - \frac{1}{4\,\Gamma_2^2} \right].
\end{equation}
The set of Eq.~(\ref{eq:Zero})-(\ref{eq:frac})-(\ref{eq:Gamma1})
is equivalent to
Eq.~(\ref{eq:CsvPart})-(\ref{eq:CsvEnerg})-(\ref{eq:CsvImpuls}) if
the parameters $\delta_1$, $1/\sigma_1$ and $1/\Gamma_1^2$ are
small.

Recall that the speed of the shocked plasma cannot exceed the speed of
the fast magnetosonic wave having a Lorentz factor~$\Gamma_{\rm
  fms}=\sqrt{\sigma_1 / ( 1 - c_{\rm s}^2 )}$ where $c_{\rm s}$ is the
sound speed. For an ultra-relativistic gas, $c_{\rm s} \approx 1 /
\sqrt{3}$ and therefore $\Gamma_2 \le \Gamma_{\rm fms} \approx
\sqrt{3\,\sigma_1/2}$.  Thus, for a super-magnetosonic upstream flow
satisfying~$\Gamma_1 \gg \sqrt{\sigma_1}$, the downstream Lorentz
factor always satisfies $\Gamma_2 \ll \Gamma_1$.  Assuming also that
upstream of the shock, contribution of the hot plasma in the sheet to
the total plasma energy is small such that~$\delta_1 \ll 1/\sigma_1$,
one reduces Eqs.~(\ref{eq:frac}) and (\ref{eq:Gamma1}) to
\begin{equation}
  \label{eq:fraclimit}
  \frac{1}{2\,\sigma_1} - \frac{1}{2\,\Gamma_2^2} =
  \delta_2 \, ( 3 - \frac{2}{Z_2} );
\end{equation}
\begin{equation}
  \label{eq:Gamma2}
  \delta_2 + \frac{1}{4\,\sigma_1} = \frac{1}{4\,\Gamma_2^2}.
\end{equation}
Combining Eqs.~(\ref{eq:Gamma2}) and (\ref{eq:fraclimit}) one
immediately gets $Z_2=2/5$.

The two remained unknowns, $\delta_2$ and $\Gamma_2$, could be found
from Eq.~(\ref{eq:Gamma2}) and the closure condition
Eq.~(\ref{eq:del2}).  In the limit $\sigma_1\gg1$, $\Gamma_{1,2}\gg 1$
the last is written, with account of $Z_2=2/5$, as
\begin{equation}
  \label{eq:Delta2}
  \delta_2 = \xi \, \frac{\beta_1}{5 \, \Gamma_2} \,
  \frac{\sigma_1 \, r_{0}}{l_1}.
\end{equation}
Making use of Eq.~(\ref{eq:Gamma2}) and Eq.~(\ref{eq:Delta2}), we can
now analyze dissipation of the alternating magnetic fields at the
shock front.

An interesting quantity is the ratio of the downstream magnetic energy
flux to the downstream matter energy flux. This is the true
magnetisation parameter of the shocked flow defined as
\begin{equation}
  \label{eq:Ratio}
  \sigma_2 =\frac{(1-\delta_2)B^2_2}
  {4\,\mu_0 [\delta_2 p_2+(1-\delta_2)p_{c2}]\Gamma_2^2}
 =\frac{2 \, ( 1 - \delta_2 ) \, n_{\rm c2} \, \sigma_1}
  {n_{\rm c1} + 4 \, \delta_2 \, n_{\rm c2} \, \sigma_1}
\end{equation}
If the dissipated fraction of the magnetic field is sufficiently
small, the flow should satisfy the ideal MHD jump conditions; then the
downstream magnetisation is given by $\sigma_2 = 2 \, \sigma_1$
(appendix~\ref{sec:AnnexeA}). Inspecting Eq.~(\ref{eq:Ratio}), this
condition leads to
  \begin{equation}
  \label{eq:D2Dissip}
  \delta_2 \ll \frac{1}{4 \, \sigma_1}
\end{equation}
Eq.~(\ref{eq:D2Dissip}) means that contribution of the plasma in the
current sheets to the plasma energy in the downstream flow remains
negligible small so that magnetic dissipation does not affect the
dynamics of the flow.  Moreover at the condition
Eq.~(\ref{eq:D2Dissip}), Eq.~(\ref{eq:Gamma2}) yields $\Gamma_2 =
\sqrt{\sigma_1}$, which is the usual result for ideal perpendicular
MHD shock.  Substituting this result to Eq.~(\ref{eq:Delta2}) gives
\begin{equation}
  \delta_2 = \xi \, \frac{1}{5} \, \frac{\sqrt{\sigma_1} \, r_{0}}{l_1}.
\end{equation}
Now the condition (\ref{eq:D2Dissip}) for the negligible dissipation
can finally be written in terms of the upstream parameters as
\begin{equation}
  \label{eq:magom}
  \sigma_1 \ll \left( \frac{5 \, l_1}{4 \, \xi \, r_{0}} \right)^{2/3}
\end{equation}

Now let a significant fraction of magnetic energy dissipate. Then the
condition opposite to Eq.~(\ref{eq:D2Dissip}) is fulfilled so that one
can neglect the second term in the left hand side of
Eq.~(\ref{eq:Gamma2}).  Together with Eq.~(\ref{eq:Delta2}) this yields
\begin{equation}
  \label{eq:Gamma2Partial}
  \frac{\Gamma_2}{\beta_2} = \frac{5}{4\,\xi} \, \frac{l_1}{\sigma_1 \, r_{0}};
\end{equation}
\begin{equation}
  \label{eq:Del2Partial}
  \delta_2 = \left(\frac{2\, \xi \, \sigma_1 \, r_{0}}{5 \,
  l_1}\right)^2.
\end{equation}
At the condition opposite to Eq.~(\ref{eq:D2Dissip}),
Eq.~(\ref{eq:Gamma2Partial}) gives $\Gamma_2<\sqrt{\sigma_1}$.  In
order to have physically meaningful results, we assume that
$\delta_2<1$, i.e. that the current sheet thickness can never exceed
the distance between two stripes. Full dissipation occurs when the
width of the current sheets becomes as large as the distance between
two successive sheets, which means $\delta_2 \to 1$.  This occurs when
the expression in the curly brackets in Eq.~(\ref{eq:Del2Partial})
goes to unity.  Now the condition of full magnetic dissipation could
be written as
\begin{equation}
  \label{eq:FullDissip}
  \sigma_1 > \frac{5 \,  l_1}{2 \, \xi \, r_{0}}
\end{equation}

When this limit is reached, the Lorentz factor downstream,
Eq.~(\ref{eq:Gamma2Partial}), becomes non-relativistic $\Gamma_2
\approx 1$. More precisely, setting $\delta_2=1$ in the exact jump
conditions~Eq.(\ref{eq:CsvEnerg}) and (\ref{eq:CsvImpuls}), we get
\begin{eqnarray}
  \label{eq:CsvEnergd2}
  2 \, \beta_1 \, n_{\rm c1} \, (1 + \sigma_1) & = &
  \beta_2 \, n_{\rm c2} \,
  \left( 1 + 4 \, \frac{n_{c2}}{n_{c1}} \, \sigma_1 \right) \\
  \label{eq:CsImpulsd2}
  n_{\rm c1} \, \left( \Gamma_1 \, \beta_1^2 \, (1 + \sigma_1) +
    \frac{\sigma_1}{2\,\Gamma_1} \right) & = &
  n_{\rm c2} \, \left\{ \Gamma_2 \, \beta_2^2 \,  \left[
      1 + 4 \, \frac{n_{\rm c2}}{n_{\rm c1}} \, \sigma_1
    \right] + \frac{1}{4\,\Gamma_2} +
    \frac{n_{\rm c2} \, \sigma_1}{n_{\rm c1} \, \Gamma_2} \right\}
  \, \frac{1}{2} \, \frac{\Gamma_1}{\Gamma_2}
  \end{eqnarray}
  In the highly magnetised and ultrarelativistic limit, we find
  \begin{eqnarray}
  \label{eq:Beta2D2}
  \beta_2 + \frac{1}{4 \, \Gamma_2^2 \, \beta_2} & = & 1
\end{eqnarray}
The only physically acceptable solution is $\beta_2=1/3$. Therefore,
in case of full dissipation, the downstream parameters are the same as
in the non-magnetized shock.

We summarise the aforementioned results in Table~\ref{tab:Critere}.
\begin{table}[htbp]
  \centering
  \begin{tabular}{|c|c|c|c|}
    \hline
    &
    $\sigma_1 \gg 5 \, l_1 / ( \xi \, r_{0}) $ &
    $ 1/5 \ll l_1 / ( \xi \, r_{0} \, \sigma_1 ) \ll \sqrt{\sigma_1}$ &
    $\sigma_1 \ll \left[ 5 \, l_1 / ( 4 \, \xi \, r_{0} ) \right]^{2/3}$ \\
    \hline
    \hline
    Reconnection level & Full & Partial & Negligible \\
    $\delta_2$  & $\approx1$ & $( 2 \, \xi \, \sigma_1 \, r_{0} / 5 \, l_1 )^2$
    & $\ll 1 / 4\,\sigma_1$ \\
    $\Gamma_2$  & $\approx 3 /  ( 2\sqrt{2}) $ &
    $ 5 \, l_1 / ( 4 \, \xi \, \sigma_1 \, r_{0} ) $ & $\sqrt{\sigma_1}$ \\
    \hline
  \end{tabular}
  \caption{Summary of the analytical criterion
    for the dissipation of alternating fields at the shock. }
  \label{tab:Critere}
\end{table}

\subsection{Numerical solution to the jump conditions}

In order to investigate the reconnection properties in less restricted
limits than those used to obtain an analytical solution to the jump
condition in Table~\ref{tab:Critere}, we solve numerically the average
conservation laws of particles, energy and momentum,
Eq.(\ref{eq:CsvPart}), (\ref{eq:CsvEnerg}), (\ref{eq:CsvImpuls})
supplemented with Eq.~(\ref{eq:del2}) for the current sheet thickness.
We remove the assumption $\sigma_1\gg1$ but keep a highly relativistic
supermagnetosonic flow, $\Gamma_1 \gg \sqrt{\sigma_1}$, with thin
current sheets, $\delta_1\ll1/\sigma_1$.

We solve the full system of Eq.(\ref{eq:CsvPart}),
(\ref{eq:CsvEnerg}), (\ref{eq:CsvImpuls}), (\ref{eq:Taval}) and
(\ref{eq:Del2}) with upstream flow conditions given by
$\Gamma_1=10^4$, $n_{\rm c1} = n_{\rm h1} = 1$, $\delta_1=10^{-8}$ and
different values for the parameter $\xi=10,10^2,10^3$.  Note that if
the flow is super-magnetosonic, $\Gamma_1 \gg \sqrt{\sigma_1}$, and
the contribution of the hot plasma in the sheets to the overall plasma
energy is negligible, $\delta_1 \ll 1/\sigma_1$, the results are
independent of $\delta_1$, $Z_1$ and $\Gamma_1$; the only important
quantity being the upstream magnetisation~$\sigma_1$.  The free
parameter~$\xi$ will be determined later by performing PIC
simulations, see Sect.~\ref{sec:Pic}.

\begin{figure}[htbp]
  \centering \includegraphics[scale=1]{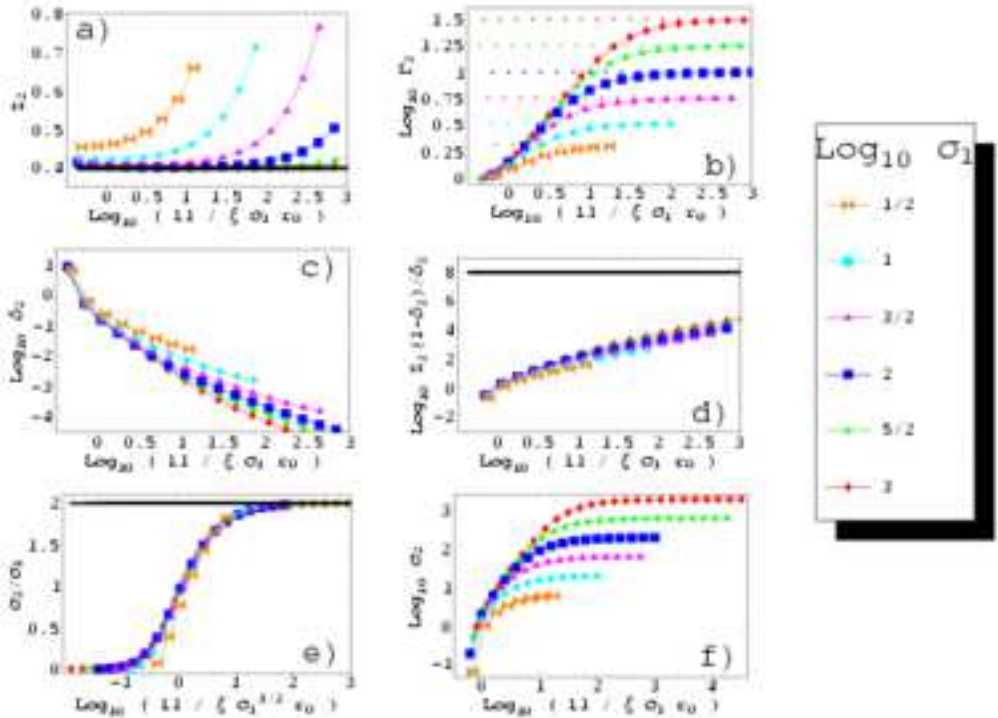}
  \caption{Several downstream parameters obtained by numerical solution
    of the jump conditions. The ratio $Z_2=n_{\rm c2}/n_{\rm h2}$ is
    shown in Fig~a), the downstream Lorentz factor in Fig~b), the
    downstream relative current sheet thickness in Fig~c).  The
    fraction of cold to hot plasma component after shock crossing,
    evaluated by Eq.~(\ref{eq:FractionD}) is shown in Fig~d). Even in
    the low dissipation limit, it is not equal to the upstream
    fraction, denoted by solid thick line at $10^8$, see
    Eq.~(\ref{eq:FractionU}).  The ratio of downstream to upstream
    magnetisation~$\sigma_2 / \sigma_1$ is shown in Fig~e), the
    downstream magnetisation in Fig~f). Note that for the
    ratio~$\sigma_2 / \sigma_1$, the abscissa scales like
    $\sigma^{-3/2}$ in order to overlap all curves independently of
    $\sigma_1$ and to match condition~(\ref{eq:magom}). All quantities
    except for fig~e) are plotted versus the adimensional
    parameter~$l_1/\xi\,\sigma_1\,r_{0}$.  Different colors associated
    with different symbols represent different $\log_{10}\,\sigma_1$
    as detailed in the legend. The asymptotes in the limit of no
    dissipation (ideal MHD shock) in fig.~b) are shown in colored
    dotted lines.}
  \label{fig:Jump1}
\end{figure}

Results for different upstream magnetisation~$\sigma_1$ and different
half wavelength of the striped wind~$l_1$ are summarised in
Fig.~\ref{fig:Jump1}. The results scale with~$\xi$ as expected so we
divide the abscissa by the parameter~$\xi$ to get ``universal''
curves.  When the contribution of the hot plasma in the current sheet
to the overall plasma energy downstream is small, i.e. for $\delta_2
\, \sigma_1 \ll 1$, dissipation is negligible and the ideal MHD jump
conditions apply.  Inspecting Fig.~\ref{fig:Jump1}e), we conclude that
this should happen whenever $\sigma_1 \le ( l_1 / \xi \, r_{0}
)^{2/3}$ so that we recover the no dissipation criterion
Eq.~(\ref{eq:magom}).  Therefore, the Lorentz factor is equal to
$\Gamma_2^2 \approx \sigma_1 + 9/8$ (\cite{1984ApJ...283..694K},
appendix~\ref{sec:AnnexeA}) as readily seen in Fig.~\ref{fig:Jump1}b).
Because the upstream flow is not necessarily highly magnetised, in
order to achieve satisfactory accuracy, we kept two terms in the
expansion of the Lorentz factor~$\Gamma_2$ with respect to $\sigma_1$
as given in \cite{1984ApJ...283..694K}. In this region, the density of
the cold part is close to the second order approximation $n_{\rm c2} /
n_{\rm c1} \approx (1 + 1/2\,\sigma_1 - 3/16\,\sigma_1^2)$.  The wind
remains entirely dominated by the electromagnetic energy flux,
$\sigma_2 \approx 2\,\sigma_1$.
The dissipation is complete when $\delta_2$ approaches unity,
Fig.~\ref{fig:Jump1}c), so that the striped structure is removed.
Equivalently, it corresponds to the case where $\sigma_2 \ll 1$.
Inspecting the Fig.~\ref{fig:Jump1}f), we expect this to happen when
$\sigma_1 \ge l_1 / \xi \, r_{0}$; so we retrieve the condition
Eq.~(\ref{eq:FullDissip}) for full Poynting flux dissipation. In this
case, the magnetic field is converted into particle heating.  The
thermalisation is complete and the flow downstream is purely
hydrodynamical.  The total energy flux is entirely carried by the
matter energy flux,~$\sigma_2 \ll 1$. Note that the region where
$\delta_2>1$ is meaningless. This can imply some negative
magnetisation for low $\sigma_1$ as clearly seen in
Fig.~\ref{fig:Jump1}e). The fraction of cold to hot particle
densities~$Z_2$ is close to $2/5$, fig.\ref{fig:Jump1}~a).  Actually,
for sufficiently large magnetisation~$\sigma_1$, it does not vary much
and remains close to this value, let reconnection be or not.  The
Lorentz factor becomes close to unity~$\Gamma_2 \approx 1$,
fig.\ref{fig:Jump1}~b).

To resume, the numerical solution of the jump conditions confirm our
analytical expectations summarised in Table~\ref{tab:Critere}.
Results have been extended to less restricted upstream flows, the
magnetisation being not necessarily very high. A good estimate for
$Z_2$ is $2/5$. It is a good approximation whenever $\sigma_1\ge100$,
as depicted by a solid thick line in Fig.~\ref{fig:Jump1}a).
 
In Fig.~\ref{fig:Jump1}d), we plot the fraction of cold to hot plasma
component given by
\begin{equation}
  \label{eq:FractionD}
  Z_2 \, \frac{1 - \delta_2}{\delta_2}
\end{equation}
It is clearly seen that this ratio is not constant and much less
than the upstream fraction given by
\begin{equation}
  \label{eq:FractionU}
  Z_1 \, \frac{1 - \delta_1}{\delta_1} = 10^8
\end{equation}
and shown by a solid thick line.  The reason is that some magnetic
dissipation occurs even at the condition (\ref{eq:magom}).  The amount
of the energy dissipated is not sufficient to affect the jump
conditions but some particles from the cold part diffuse across the
magnetic field lines and enter the sheets.

In a final step, we determine the free parameter~$\xi$ with help of
PIC simulations.  This is discussed in the next section.

\section{TERMINATION SHOCK: PIC SIMULATIONS}
\label{sec:Pic}

We designed a fully relativistic and electromagnetic 1D PIC code
following the algorithms described in \cite{Birdsall2005}. Particle
trajectories are advanced in time by integrating the relativistic
equation of motion due to the Lorentz force.  The longitudinal
electric field is found by solving Poisson equation whereas the
transverse component of the electromagnetic field are computed by the
remaining component of Maxwell equations introducing a left and
right-going wave such as $F^\pm = E_{\rm y} \pm c \, B_{\rm z}$, see
\cite{Birdsall2005}. The simulation is one dimensional in space along
the $x$-axis and two dimensional in velocity in the plane $(xOy)$.

The striped wind propagates along the $x$-axis and hits a wall located
at the right boundary of the simulation box of length~$L$.  We also
impose no incoming electromagnetic wave at this boundary.  The
particle momentum vector possesses a longitudinal component~$p_{\rm
  x}$ and a transversal component~$p_{\rm y}$.  They propagate from
the left to the right and hit the solid wall at $x=L$. This means that
particles are reflected at this boundary, the $p_{\rm x}$ momentum is
reversed whereas the $p_{\rm y}$ component remains unchanged. The
magnetic field is directed along the $z$-axis and reverses polarity
when crossing each current sheet.  Electromagnetic waves leave the
simulation box at the right boundary without any reflection.  Similar
simulations have already been performed by \cite{2005AdSpR..35.1112L}.

\begin{figure}[htbp]
  \centering
  \begin{tabular}{cc}
    \includegraphics[scale=0.9]{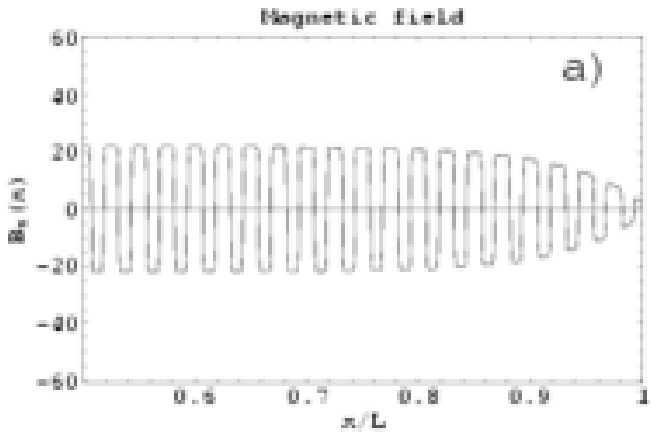} &
    \includegraphics[scale=0.9]{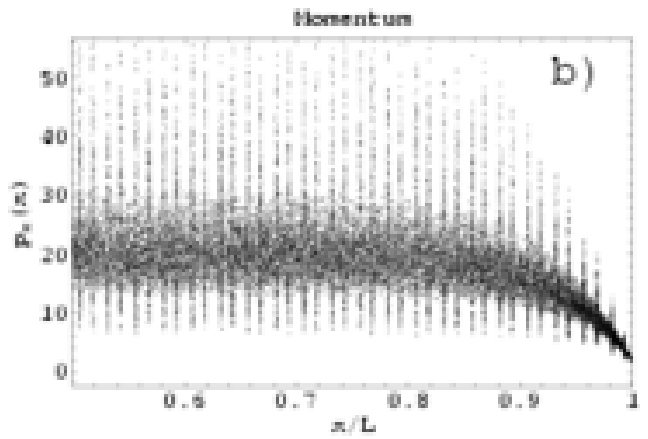} \\
    \includegraphics[scale=0.9]{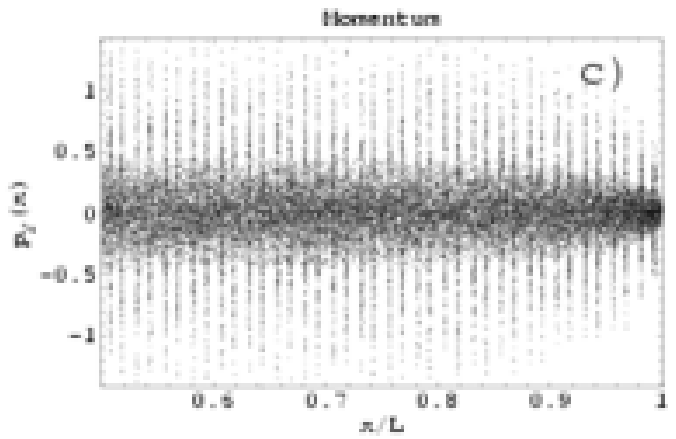} &
    \includegraphics[scale=0.9]{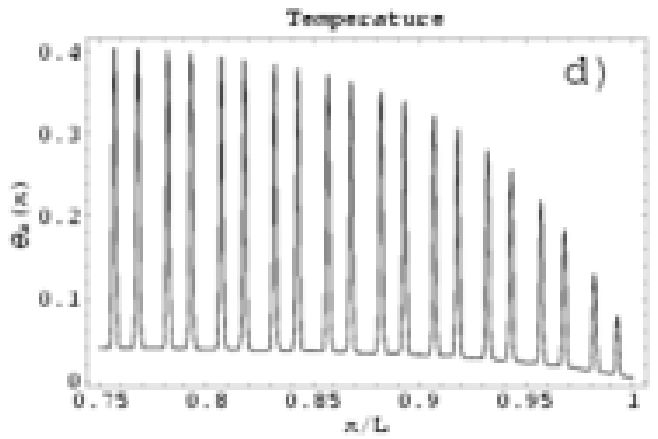} \\
    \includegraphics[scale=0.9]{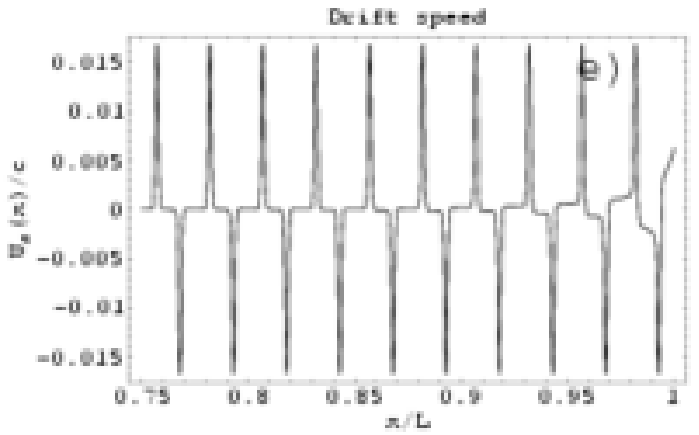}
  \end{tabular}
  \caption{Initial conditions. The geometry of the striped
    magnetic field is shown in Fig.~a). The longitudinal~$p_{\rm x}$
    and transversal~$p_{\rm y}$ component of the momentum of one
    species are seen in Fig.~b) and~c) respectively. The temperature
    and the drift speed for one species (the speed of the other
    species is equal and opposite) are shown respectively in Fig.~d)
    and e).}
  \label{fig:Initial}
\end{figure}

A typical initial situation showing the magnetic field structure, the
lepton distribution functions, the temperature and the drift speed is
shown in Fig.~\ref{fig:Initial}. Close to the right wall, we impose a
decreasing magnetic field strength~$B_{\rm z}(x)$ which almost
vanishes at $x=L$, Fig.~\ref{fig:Initial}a).  The magnetic polarity
reversal is smooth, it is not squared but evolves according to the
following expression
\begin{equation}
  \label{eq:Binit}
  B_{\rm z}(x) = B_0 \, \tanh[\Delta \, ( a + b \, \cos( \lambda \, x ) )] \, D(x)
\end{equation}
where $B_0$ is the maximal intensity of the magnetic field, $a$, $b$,
$\lambda$ and $\Delta$ are constants, $\Delta$ prescribing the
thickness of a current sheet and $\lambda$ the number of periods in
the simulation box. This particular expression is dictated by the
structure of the striped wind, see for instance Eq.~(1) in
\cite{2005ApJ...627L..37P}. $D(x)$ is a function introduced to
decrease the magnetic field intensity close to the wall, at $x=L$.
Moreover, because magnetic pressure is balanced by gaseous pressure,
the temperature~$T(x)$ in the gas also decreases close to the right
wall where it almost vanishes, Fig.~\ref{fig:Initial}d), in accordance
with the magnetic field behavior. Lower temperature implies weaker
spread in particle momentum space, Fig.~\ref{fig:Initial}b) and c)
(note that we keep track of only 50.000~particles, chosen randomly at
each time step, in order to avoid too large data files). We also take
the bulk Lorentz factor of the flow, ~$\Gamma_1$, decreasing near the
right wall. The average longitudinal momentum~$p_{\rm x}$ also
decreases in relation with the decreasing $\Gamma_1$.  With this
initial condition, we avoid the formation of too large a transient
when the plasma first collides with the right wall.

The plasma moves in the positive $x$ direction with the bulk Lorentz
factor $\Gamma_1=20$. The temperature in the plasma is obtained from
the equilibrium conditions, namely pressure balance between gaseous
and magnetic part. In the current sheet, the temperature is much
higher than in the magnetised part, Fig.~\ref{fig:Initial}d).  This
implies a much larger spread in momentum as seen in
Fig.~\ref{fig:Initial}b) and c).  The average magnetic field
$\alpha=<B>/B_0$ is not necessarily zero because two successive
stripes generally have different width but the same magnetic field
intensity~$B_0$.  In the pulsar wind, the average magnetic field only
vanishes in the equatorial plane, $\theta=\pi/2$ in spherical
coordinates~$(r, \theta, \varphi)$. Actually, in
Fig.~\ref{fig:Initial}a) as well as in the simulation results
presented below (Fig.~\ref{fig:PIC2}a) and Fig.~\ref{fig:PIC1}a)), the
average magnetic field is $\alpha \approx 0.12$. We choose an average
value different from zero not because of numerical stability
requirement but because we want to demonstrate that even in case of
full dissipation the average magnetic flux downstream the shock is
preserved.  We also performed a set of simulations with zero average
magnetic field. The results obtained are not affected and remain
qualitatively the same in both cases as will be shown later.
Moreover, passing from one current sheet to the next one, the drift
speed for each species reverses sign, Fig.~\ref{fig:Initial}e).
Indeed, in a current sheet, the magnetic field, oriented in the
$z$~direction, is sustained by an electric current flowing in the
$y$~direction.  This current is generated by electrons moving in,
let's say, the positive $y$~direction at a speed $U_s$ whereas
positrons are moving in the opposite direction at a speed $-U_s$.
Charge neutrality is maintained but a net total current exists.

For the results presented in this section, the simulation box is
divided into $2\times10^5$~cells and we used $8\times10^5$ particles
for each species, i.e. $8\times10^5$ electrons and $8\times10^5$
positrons. Therefore, on average, there are only $4+4$~particles per
cell. Although this number seems rather small, we checked that
increasing the number of particles per cell do not improve the
accuracy of our results. Thus, we performed the whole set of
simulations with this resolution. To support our statement, we will
show examples with $40+40$~particles per cell.  The particle number
density in the hot unmagnetised part, i.e. in the sheets, is five
times higher than in the cold magnetised part, $n_{\rm h1} = 5\,n_{\rm
  c1}$.  The time step is chosen such that $\omega_B \, \Delta t =
0.5$.  The resulting upstream Larmor radius, deduced from this value
and from $\Gamma_1=20$ according to Eq.~(\ref{eq:Larmor1}), is equal
to~$r_{0} = 40$~cells.  Half a wavelength of the striped wind is $l_1
= 2500$~cells and the relative current sheet thickness is~$\delta_1 =
0.1$. At the left boundary of the simulation box, no incoming plasma
is injected.  The upstream plasma is simply flowing to the right and
leaves an empty space behind it with a constant magnetic field
advected at the flow bulk speed~$\Gamma_1$.  In order to keep
meaningful results, the simulation has to be stopped at the time
$t_{\rm f} = L / 2c$ before the electromagnetic wave starting at $t=0$
at the right wall reaches this vacuum region, propagating from left to
right starting from~$x=0$ at $t=0$.

In all runs, the simulation box starts with a inhomogeneous plasma
made of electrons and positrons following a 2D relativistic Maxwellian
distribution function in their proper frame given by
\begin{equation}
  \label{eq:fdd2drelat}
  f_s(x,p_{\rm x},p_{\rm y}) = \frac{N_s}{2\,\pi} \, \frac{e^{- ( \gamma - 1 ) / \Theta_s}}
  {m_s^2 \, c^2 \, \Theta_s \, ( 1 + \Theta_s )}
\end{equation}
with relativistic temperature $\Theta_s = k_{\rm B} \, T_s / m_s \,
c^2$ and Lorentz factor $\gamma = \sqrt{1 + (p_{\rm x}^2 + p_{\rm
    y}^2) / m_s^2 \, c^2}$. The temperature~$\Theta_s$ is obtained
from the pressure equilibrium condition Eq.~(\ref{eq:EqPressAmont})
and shown in Fig.~\ref{fig:Initial}d).  $N_s$ is the particle number
density in the proper frame, $m_s$ the mass of a particle of each
species.  Because they are equal for electrons and positrons, we
note~$m_s=m$.  We have to distinguish between 3 reference frames
\begin{itemize}
\item $R_\mathrm{s}$: the species frame or proper frame of reference
  for each species, electrons and positrons, in which the distribution
  function is given by the 2D relativistic Maxwellian
  Eq.~(\ref{eq:fdd2drelat})~;
\item $R_\mathrm{w}$: the wind frame of reference in which the leptons
  are counter-streaming with velocity $\pm U_s$~;
\item $R_0$: the observer frame or simulation box frame or lab frame
  of reference in which the wind (the current sheets) is propagating
  in the positive $x$-direction with a Lorentz factor~$\Gamma_1$.
\end{itemize}
To upload the initial distribution function of each species in the lab
frame, we have to make two successive Lorentz transformations, the
first one leading from $R_\mathrm{s}$ to $R_\mathrm{w}$ and the second
one leading from $R_\mathrm{w}$ to $R_0$.

We performed many runs with different initial Lorentz factor of the
wind and different magnetisation.  We give a typical sample of runs
demonstrating full dissipation of the magnetic field or no dissipation
at all.

Because the velocity space is only two-dimensional in our simulations,
see the distribution function Eq.~(\ref{eq:fdd2drelat}), the adiabatic
index for the relativistic plasma is $\gamma=3/2$ instead of the usual
$4/3$. This different index will affect the parameters downstream the
shock like the temperature, the Lorentz factor and the magnetisation.
The general jump conditions for arbitrary adiabatic index are derived
in appendix~\ref{sec:AnnexeA}. These formulae will help us to make
quantitative comparisons between the analytical results presented
before (and made for $\gamma=4/3$) and the PIC simulations.

\subsection{Negligible dissipation}

Here we show details of a typical example where magnetic dissipation
is negligibly small.  The magnetisation between the sheets is set
to~$\sigma = 3$. It corresponds to the definition of $\sigma_1$ in
Eq.~(\ref{eq:Magnetisation1}).  The true
magnetisation~Eq.~(\ref{eq:SigmaMoy}), equal to $\sigma \approx 2.7$,
differs from the one in the cold part~$\sigma$ because in our PIC
simulations, we need to resolve the structure of the sheet (sheet
thickness larger than the Larmor radius) therefore the contribution
from the cold unmagnetised part is not negligible.  The true average
magnetisation is actually roughly 10\% lower than the one in the cold
part due to the relative current sheets thickness $\delta_1 = 0.1$.
The true magnetisation, averaged over one wavelength of the striped
wind , is given in the rest frame of the downstream plasma by
\begin{equation}
  \label{eq:SigmaMoy}
  \langle \sigma \rangle =
  \frac{\langle \mathrm{electromagnetic \;\; energy \;\; density} \rangle}
  {\langle  \mathrm{enthalphy \;\; density} \rangle}
\end{equation}
For a fluid with adiabatic index $\gamma=3/2$, the average enthalpy
density (in the rest frame) is given by
\begin{equation}
  \label{eq:Enthalpie}
  \langle \mathrm{enthalphy \;\; density} \rangle = 3/2 \, \langle
  \mathrm{internal \;\; energy \;\; density}\rangle - 1/2 \,
  \langle\mathrm{particle \;\; number \;\; density}\rangle \, c^2
\end{equation}
The internal energy includes the rest mass energy as well as the
kinetic energy of the particles.  This formula is only valid in the
rest frame of the fluid. In our computations, the frame of the
simulation box corresponds to the rest frame of the downstream plasma,
therefore Eq.~(\ref{eq:Enthalpie}) applies to this plasma but not to
the upstream one which moves with a Lorentz factor~$\Gamma_1$ with
respect to the simulation box. Thus, we have first to transform back
to the rest frame of the upstream plasma and then apply
Eq.~(\ref{eq:SigmaMoy}) and (\ref{eq:Enthalpie}). The upstream rest
frame is easily found because we impose the Lorentz factor of the
upstream plasma with respect to the simulation box which is also the
rest frame of the downstream plasma. Therefore, the Lorentz transform
in the upstream frame involves simply~$\Gamma_1$. Note however that
this is correct only well upstream and not in the shock itself. The
bulk Lorentz factor differs from $\Gamma_1$ whenever a precursor
arrives. Nevertheless, we are only interested in the average
quantities when the front shock has passed through the plasma. The
precise values within the shock front are not significant for our
study. Note that the {\it true local} wavelength upstream as well as
downstream the shock, necessary for the averaging $<>$ in
Eq.~(\ref{eq:SigmaMoy}) and (\ref{eq:Enthalpie}), is determined by
looking for the locations where the magnetic field vanishes.  Indeed,
due to compression of the plasma after shock crossing, the length of
the stripes downstream are reduced by a factor~$\beta_1/\beta_2(x)>1$
compared to their value upstream, according to
Eq.~(\ref{eq:wavelength}). Note also that the downstream wavelength
can vary slightly (a few percent) from one period to the next one, due
to some perturbations. These perturbations propagate also in the
upstream flow, causing some small changes in the upstream wavelength
too. All these variations are taken into account when computing the
averaged quantities presented in the figures.

\begin{figure}[htbp]
  \centering
  \begin{tabular}{cc}
    \includegraphics[scale=0.9]{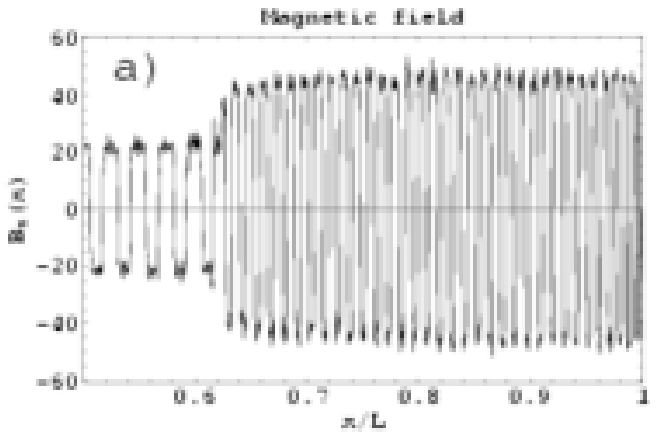} &
    \includegraphics[scale=0.9]{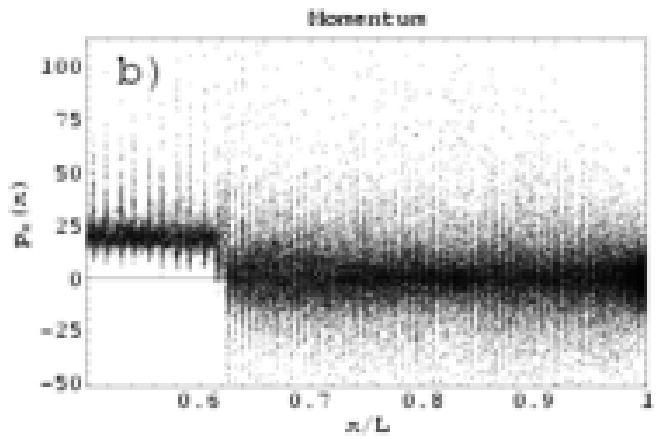} \\
    \includegraphics[scale=0.9]{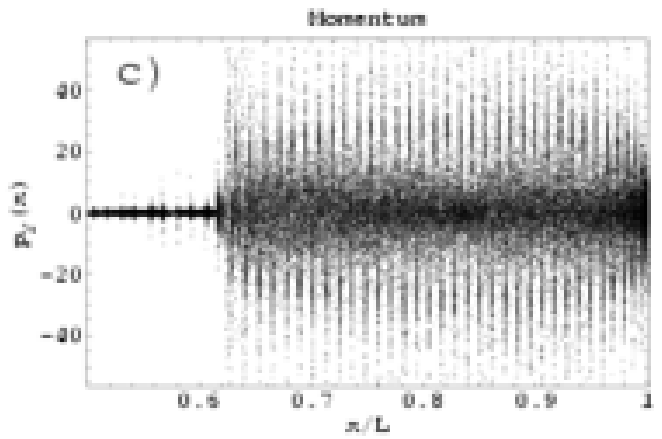} &
    \includegraphics[scale=0.9]{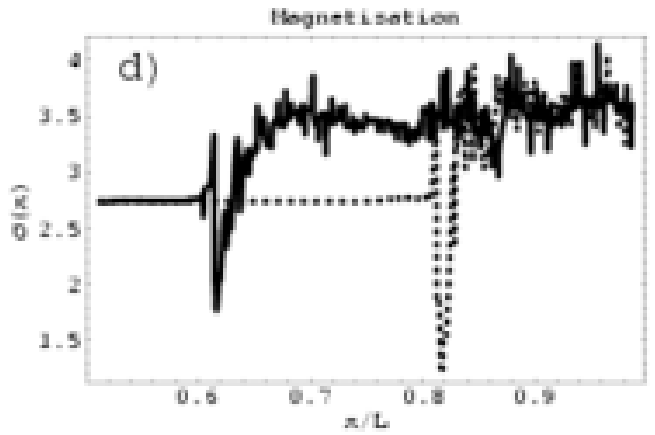} \\
    \includegraphics[scale=0.9]{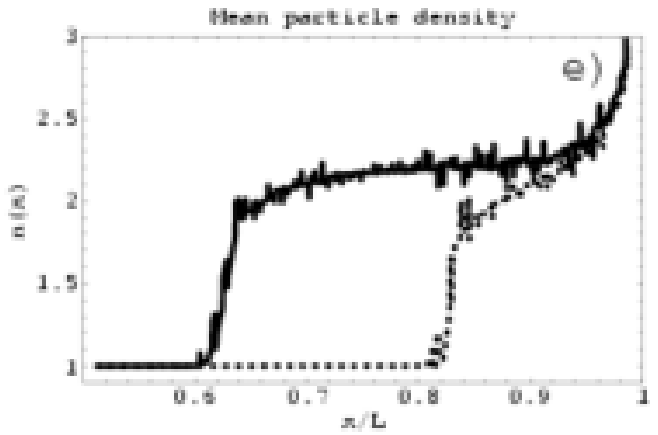}
  \end{tabular}
  \caption{Example of run showing no magnetic reconnection
    at the termination shock. The Lorentz factor of the wind is
    $\Gamma = 20$ and the magnetisation $\sigma = 3$. The magnetic
    field is shown in Fig.~a), the particle longitudinal and
    transverse momentum component in Fig.~b) and~c) respectively, the
    intermediate, $t=2.5\times10^4$ (dotted line) and final,
    $t=5\times10^4$ (solid line) downstream magnetisation in Fig.~d)
    and the intermediate (dotted line) and final (solid line) average
    particle density number downstream in Fig.~e). All quantities are
    plotted against the normalised abscissa~$x/L$.}
  \label{fig:PIC2}
\end{figure}

The results of the run are summarised in Fig.~\ref{fig:PIC2} for a
final time of simulation $t_{\rm f} = L / 2 \, c = 5 \times 10^4$.
This corresponds to the time needed by an electromagnetic wave to
propagate from the wall at the right boundary, starting at $t=0$ and
arriving at the middle of the simulation box~$L/2$ at the time~$t_{\rm
  f}$.  The shock front is very sharp with a very small thickness and
located at approximately~$x/L \approx 0.62$. It propagates in the
negative $x$ direction, starting from the wall located at the right
boundary as before.  Looking at the structure of the magnetic field,
Fig.~\ref{fig:PIC2}a), the stripes downstream are perfectly preserved.
They are still clearly recognisable in Fig.~\ref{fig:PIC2}b) and c).
They are only subject to a compression of a factor roughly equal to
two.  Thus the wavelength of the wind in the downstream frame has been
divided by two whereas the magnetic field strength, due to magnetic
flux conservation, increases by a factor two. The mean particle
density in the shocked plasma, Fig.~\ref{fig:PIC2}e), reaches a value
close to two, in accordance with the magnetic field compression ratio
of two.  Because we average over one wavelength of the striped wind,
the difference in hot and cold density is smoothed out.

The factor two can be explained as follows. The shock propagates with
velocity close to c.  In the shock frame, the magnetic field strength
and the particle number density remain nearly the same, $B_2=B_1$ and
$n_2=n_1$ (in the high $\sigma$ limit the shock is weak!).  However,
in the simulation frame, which is the downstream frame, they vary
because of Lorentz transformations. Indeed, let's note $n_{1/2}$ the
density of the upstream plasma, proper density~$n_1'$, as measured in
the rest frame of the downstream plasma. Let also $\Gamma_{1/2}$ be
the relative Lorentz factor of the incoming flow as measured in the
shocked plasma. It is easy to show that a Lorentz transformation gives
\begin{equation}
  \label{eq:Gamma1s2}
  \Gamma_{1/2} = \Gamma_1 \, \Gamma_2 \, ( 1 - \beta_1 \, \beta_2 )
\end{equation}
Thus the upstream density as measured in the downstream plasma is
\begin{equation}
  \label{eq:n1s2}
  n_{1/2} = \Gamma_{1/2} \, n_1'
\end{equation}
With help of the jump condition for particle number density
conservation, we found
\begin{equation}
  \label{eq:n2sn1s2}
  \frac{n_2'}{n_{1/2}} = \frac{\Gamma_1\,\beta_1}{\Gamma_{1/2} \, \Gamma_2\,\beta_2}
\end{equation}
In the ultra-relativistic limit, $\beta_1\to1$, we get
\begin{equation}
  \label{eq:n2sn1s22}
  \frac{n_2'}{n_{1/2}} \approx \frac{1+\beta_2}{\beta_2}
\end{equation}
Therefore the factor~2 when $\beta_2\approx1$. Actually, because the
magnetisation is not very high, $\sigma_1=3$, in the shock frame, the
downstream plasma propagates with a Lorentz factor $\Gamma_2 \approx
\sqrt{5\,\sigma_1/4}$ (see appendix~\ref{sec:AnnexeA}), corresponding
to a speed $\beta_2=0.86$.  Inserting this value in
Eq.~(\ref{eq:n2sn1s22}), we found $n_2'/n_{1/2}\approx2.15$, very
close to the average density in the shocked plasma, see
Fig.~\ref{fig:PIC2}e). Therefore, the shock front propagates in the
negative $x$-direction with the speed $\beta_2$.  Starting from $t=0$
at $x=L$, at the end of our simulation, it traveled a distance $\Delta
x = \beta_2 \, c/2*t_{\rm f} \approx 0.43\,L$.  Therefore the shock
front should be located at $x/L=0.57$.  However, the magnetisation at
the initial time close to the wall is very weak, close to unity.
Therefore, the shock speed is smaller at earlier times. That is why
the shock front is located at only $x/L \approx 0.62$ as can be
checked by inspecting Fig.~\ref{fig:PIC2}a) and b) and c) where a
sharp transition is observed between the upstream and downstream flow.

Another way to estimate the speed of the shock is to compare the
position of the shock front at the final snapshot, ($x/L \approx
0.62$) and at the penultimate snapshot ($x/L \approx 0.664$).
Moreover, the time between these two snapshots is $\Delta t = L /
20\,c$.  Therefore, the speed of the shock front is roughly $\beta_2
\approx 0.044 \, L / ( L / 20 ) \approx 0.88 $, close to the estimate
made above.

\begin{figure}[htbp]
  \centering \includegraphics[scale=0.9]{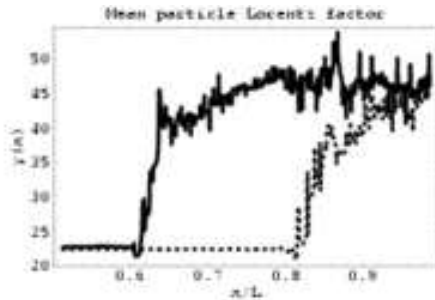}
  \caption{Mean particle Lorentz factor averaged over
    one period of the striped wind at an intermediate,
    $t=2.5\times10^4$ (dotted line) and final, $t=5\times10^4$ (solid
    line) time of simulation.}
  \label{fig:EnKin}
\end{figure}

The particles downstream are not thermalised.  Indeed, the particle
momentum which is mainly directed along the propagation axis of the
wind in the upstream flow, $p_{\rm y}\ll p_{\rm x}$, is only weakly
disturbed by the shock.  The magnitude of $p_{\rm x}$ is not altered
by the shock.  Nevertheless, after crossing the shock, both
longitudinal and transversal momenta are of the same order of
magnitude, $p_{\rm y} \approx p_{\rm x}$. The mean particle Lorentz
factor,~$\gamma$, is roughly increased by a factor of two,
Fig.~\ref{fig:EnKin}.  Although the transition for the transverse
momentum component is very sharp, the stripes are still clearly
identifiable, Fig.~\ref{fig:PIC2}b) and c).  The magnetisation
downstream is increased by a factor roughly equal to two, as expected
from the analysis of Sect.~\ref{sec:Solutions}, Fig.~\ref{fig:PIC2}d),
demonstrating that the magnetic field does not dissipate.
Contrary to the run presented in the next section, there is no
electromagnetic precursor, the flow is not perturbed by the shock
front.  The situation is very similar to the ideal relativistic MHD
shock.  The magnetic field lines, frozen into the plasma, have to
follow the motion imposed by the matter.  The shocked plasma is
compressed, and due to the frozen magnetic flux, the stripes are also
compressed in the same ratio.

\begin{figure}[htbp]
  \centering
  \begin{tabular}{cc}
    \includegraphics[scale=0.9]{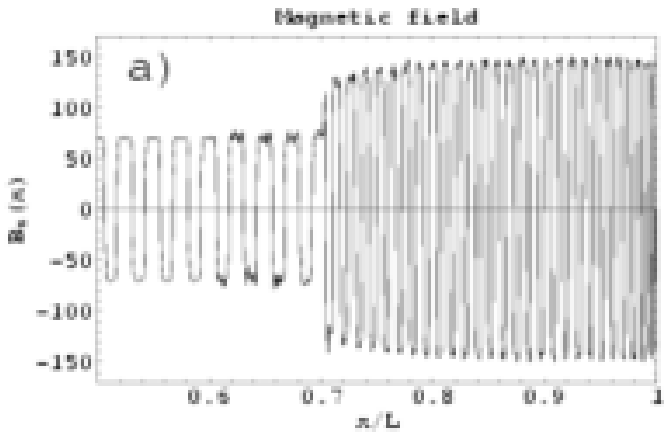} &
    \includegraphics[scale=0.9]{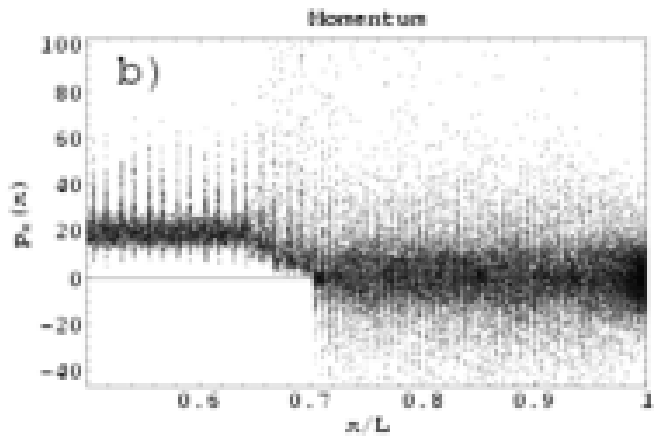} \\
    \includegraphics[scale=0.9]{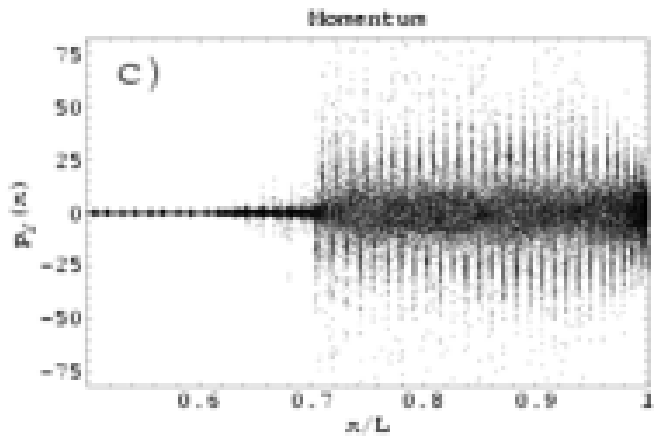} &
    \includegraphics[scale=0.9]{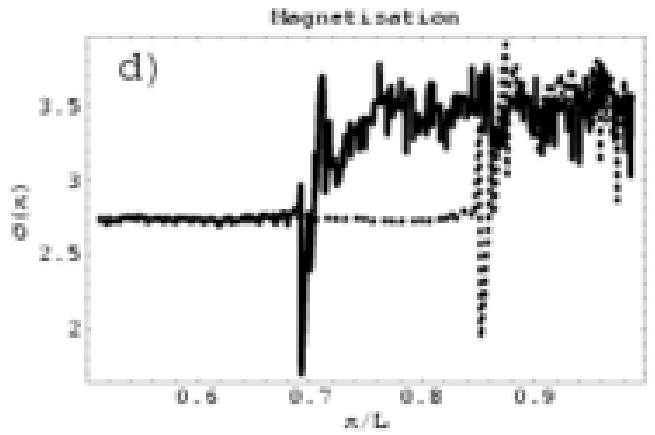} \\
    \includegraphics[scale=0.9]{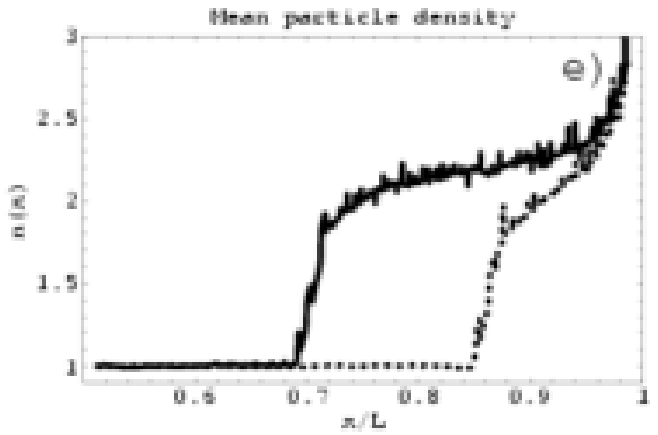}
  \end{tabular}
  \caption{Same as Fig.~\ref{fig:PIC2} but with
    ten times more particles per cell, i.e. 40 positrons and 40
    electrons per cell. The results are nearly the same as in
    Fig.~\ref{fig:PIC2}. However, we stopped the run when made sure
    that everything proceeds like the run with $4+4$~pairs so that the
    position of the shock is different.}
  \label{fig:PIC2b}
\end{figure}

Increasing the number of particles per cell will not improve the
accuracy. Indeed, we performed another run with ten times more
particles per cell ($40+40$), see Fig.~\ref{fig:PIC2b} and compare
with Fig.~\ref{fig:PIC2}. Therefore, using only $4+4$ particles per
cell is justified.

\begin{figure}[htbp]
  \centering
  \begin{tabular}{cc}
    \includegraphics[scale=0.9]{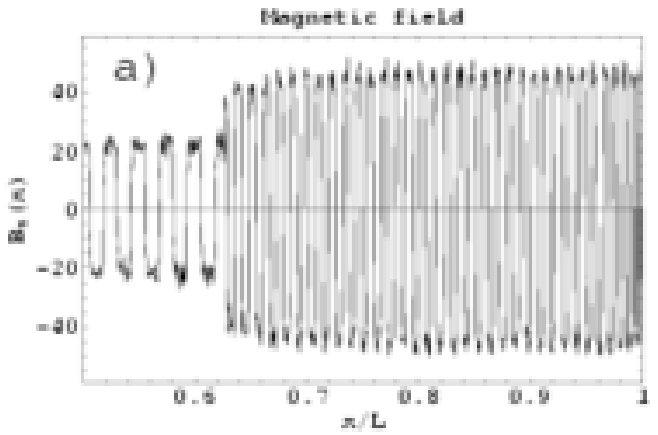} &
    \includegraphics[scale=0.9]{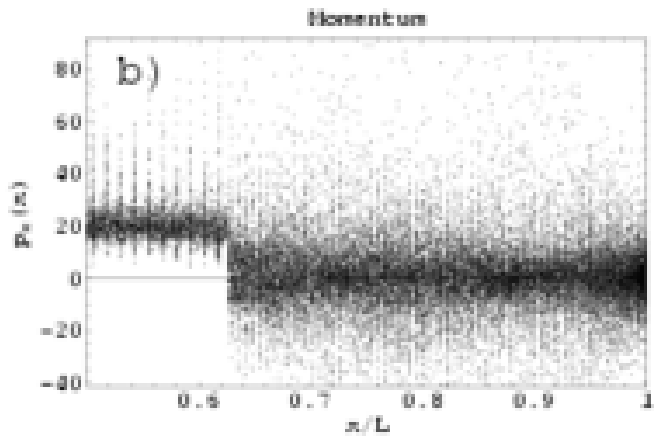} \\
    \includegraphics[scale=0.9]{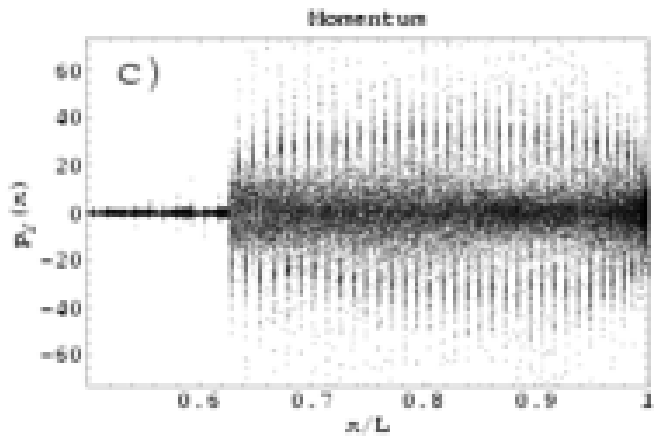} &
    \includegraphics[scale=0.9]{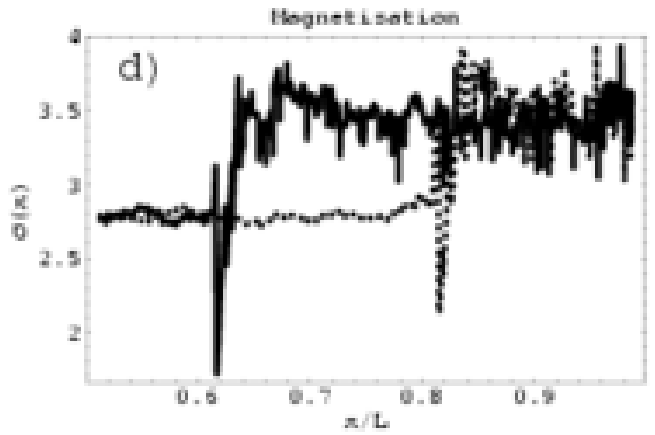} \\
    \includegraphics[scale=0.9]{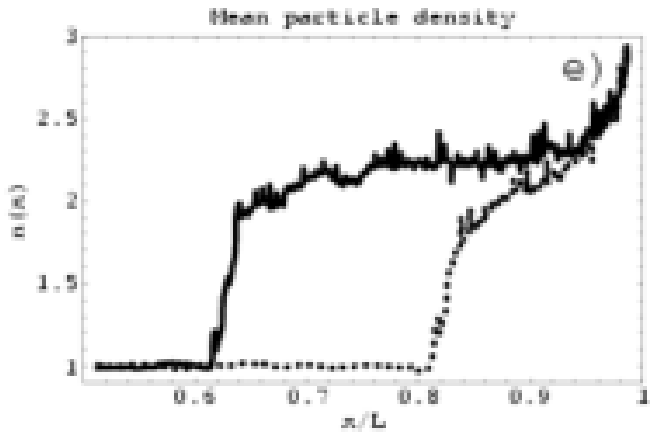}
  \end{tabular}
  \caption{Same as Fig.~\ref{fig:PIC2} but with
    zero average magnetic field, $\alpha=0$. The results are nearly
    the same as in Fig.~\ref{fig:PIC2}. }
  \label{fig:PIC2c}
\end{figure}

Finally, we show an example with zero average magnetic field
$\alpha=0$ in Fig.~\ref{fig:PIC2c} to prove that this parameter
$\alpha$ does not affect much the flow downstream provided $\alpha\ll
1$.

\subsection{Full dissipation}

In this paragraph, we discuss in detail a typical example of full
magnetic reconnection in the striped wind.  The magnetisation in the
cold magnetised part, i.e. between the sheets, is set to~$\sigma =
45$. The true average magnetisation, $\sigma_1\approx40.0$, is here
again roughly 10\% lower than the one in the cold part due to the
relative current sheets thickness $\delta_1 = 0.1$.

The only change we made compared to the previous case is to take
a higher magnetisation.  This has dramatic consequences on the flow
downstream as we will see now.

\begin{figure}[htbp]
  \centering
  \begin{tabular}{cc}
    \includegraphics[scale=0.9]{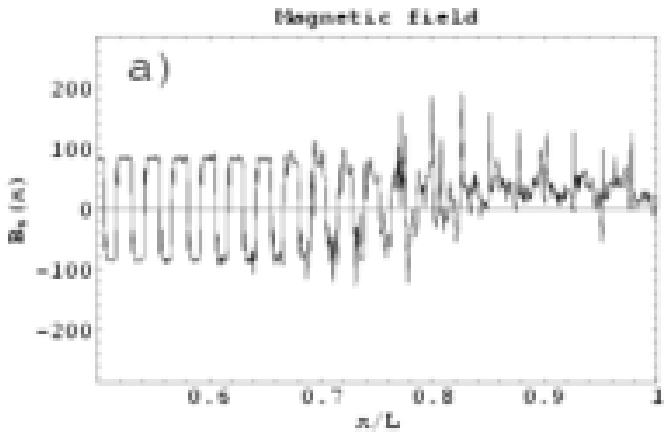} &
    \includegraphics[scale=0.9]{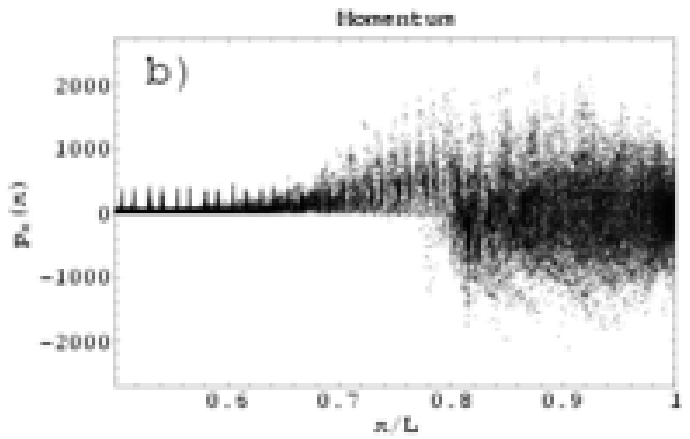} \\
    \includegraphics[scale=0.9]{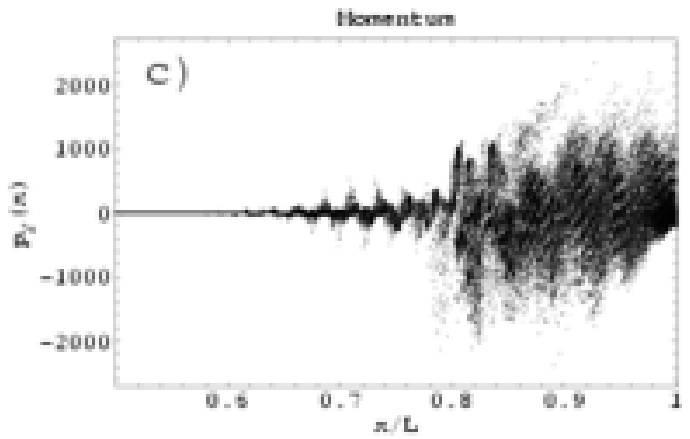} &
    \includegraphics[scale=0.9]{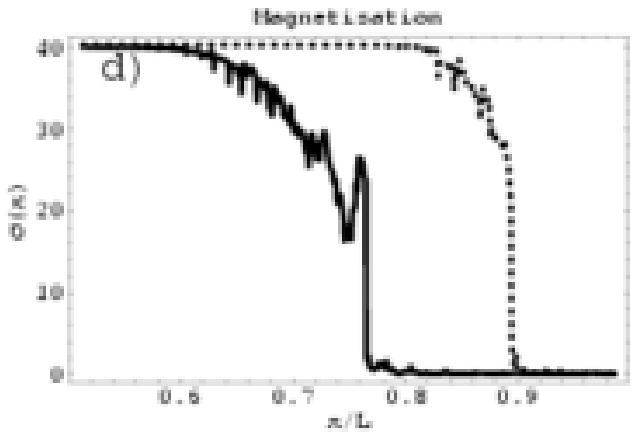} \\
    \includegraphics[scale=0.9]{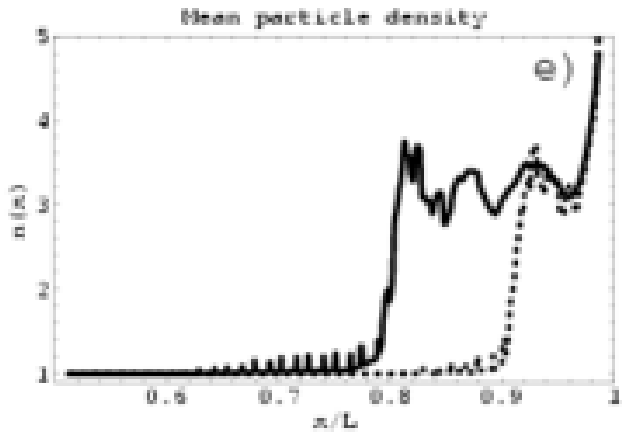}
  \end{tabular}
  \caption{Example of run showing the magnetic reconnection
    at the termination shock. The Lorentz factor of the wind is
    $\Gamma = 20$ and the magnetisation~$\sigma = 45$. The same
    quantities as in Fig.~\ref{fig:PIC2} are plotted.}
  \label{fig:PIC1}
\end{figure}

The simulation results are shown in Fig.~\ref{fig:PIC1} for the final
time $t_{\rm f} = L / 2 \, c = 5 \times 10^4$. At this snapshot, the
shock front is located at approximately~$x/L\approx0.75$ (see below
how this location was determined). It propagates in the negative $x$
direction, starting from the wall located at the right boundary~$x=L$.
Looking at the structure of the magnetic field, Fig.~\ref{fig:PIC1}a),
it is clearly seen that the stripes, after crossing the shock are
destroyed.  Nevertheless, because the average magnetic field does not
vanish, a small DC component remains, of the order of $\alpha=0.12$.
The particles downstream are fully thermalised. Indeed, in the
upstream flow corresponding to~$x/L<0.78$, the particle momentum is
mainly directed along the propagation axis of the wind, longitudinal
momentum $p_{\rm x}$ depicted in Fig.~\ref{fig:PIC1}b), with non
negligible transverse momentum~$p_{\rm y}$ only within the current
sheets, Fig.~\ref{fig:PIC1}c). The upstream longitudinal
component~$p_{\rm x}$ is by orders of magnitude larger than the
transverse component~$p_{\rm y}$, $p_{\rm y}\ll p_{\rm x}$.  The
particle momentum distribution function is randomised when crossing
the shock discontinuity, i.e. in the region defined by $x/L>0.75$, and
therefore downstream the two component of the momentum are of the same
order of magnitude, $p_{\rm y}\approx p_{\rm x}$ and by orders of
magnitude larger than before the shock. The enhancement in the
momentum of the particle in both directions, $p_{\rm x}$ and $p_{\rm
  y}$, is explained by the Poynting flux dissipation. Magnetic energy
is converted into particle kinetic energy. There are no more spikes in
the $p_{\rm x}$ graph proving that the stripes disappeared.  The
magnetisation downstream is almost zero, solid curve in
Fig.~\ref{fig:PIC1}d), indicating that the alternating magnetic field
has completely dissipated into particle heating as expected.  In that
case, the flow downstream is hydrodynamical and the compression ratio
close to three.  The mean particle density in the shocked plasma,
solid curve in Fig.~\ref{fig:PIC1}e), indeed reaches a value close to
three.

The factor three can be explained in the following way. Assuming full
dissipation of the magnetic field, the upstream flow is strongly
magnetised whereas the downstream plasma is purely hydrodynamical. For
an ultra-relativistic gas, the adiabatic index is $\gamma = 4/3$. This
is true for particles evolving in a three dimensional velocity space
in which they possess three degree of freedom in translation motion.
However, in our simulations, the distribution function,
Eq.~(\ref{eq:fdd2drelat}), is only two-dimensional in velocity space
and the corresponding adiabatic index is thus $\gamma = 3/2$. Solving
the MHD jump conditions for this 2D plasma, it is easily found that
the downstream plasma velocity in the rest frame of the shock is
$\beta_2 = 1/2$ (2D velocity space) instead of the traditional
$\beta_2 = 1/3$ (3D velocity space).  Using Eq.~(\ref{eq:n2sn1s22}),
the ratio of the particle number density expressed in the downstream
frame (simulation box frame) is $n_2'/n_{1/2} = 3$ as expected. Again,
the front shock propagates in the negative $x$-direction with the
speed $\beta_2=1/2$. Starting from $t=0$ at $x=L$, at the end of our
simulation, it traveled a distance $\Delta x=c/2\,t_{\rm f}=L/4$.
Therefore the shock front should be located at $x/L=0.75$. However,
because of some small retardation effect to initiate the shock front,
it is seen to lie at $x/L\approx0.78$ by inspecting
Fig.~\ref{fig:PIC1}a) and b) and c).

It is also interesting to note that the incoming flow is already
perturbed before entering the shock. The magnetisation gradually
decreases and becomes already very small in the region just before the
shock discontinuity. The stripes start to be significantly disturbed
already at $x/L\approx0.6$.  This phenomenon is probably due to an
electromagnetic precursor, i.e.  and electromagnetic wave generated at
the shock front and propagating downwards into the upstream plasma
\citep{1992ApJ...390..454H, 1992ApJ...391...73G}.  Moreover, from the
momentum component~$p_{\rm x}$, we conclude that the flow is heated
already well upstream of the shock.  If the strong electromagnetic
precursor would be generated immediately at the initial time $t=0$, we
would expected it to influence half of the simulation box from
$x/L=0.5$ to $x/L=1$ at the final time $t_{\rm f}$.  However, the
amplitude of the precursor is initially small because due to the
chosen initial distribution of the flow parameters
(Fig.\ref{fig:Initial}), the energy of the plasma entering the shock
is initially small. The strong shock is formed after some time (about
0.05L/c) when the highly relativistic plasma enters the shock. Thus a
strong enough precursor is generated not from the beginning of the
simulations and therefore the flow is significantly perturbed at $x/L
\gtrsim 0.55$.
The effect of a precursor was not included in our analytic analysis.
It is performed with the assumption that the dissipation is weak; in
this case, there is no significant precursor.  In any case, our
analysis is valid provided the downstream flow is settled into a
striped wind. Then we apply the conservation laws to flows well
upstream and well downstream of the shock and processes within the
``black box'' containing the shock and possible precursors do not
affect the results.

In case of high upstream magnetisation, the flow in the wind is
dominated by the dynamics of the magnetic field, i.e. gaseous pressure
negligible compared to magnetic pressure. Therefore the shocked plasma
has to reorganise in such a way to keep the striped structure.  If
this plasma is unable to maintain the current required by the magnetic
field, the latter will dissipate.

\begin{figure}[htbp]
  \centering
  \begin{tabular}{cc}
    \includegraphics[scale=0.9]{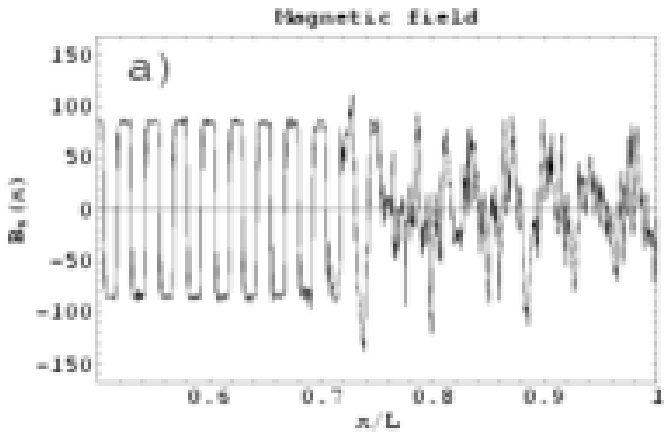} &
    \includegraphics[scale=0.9]{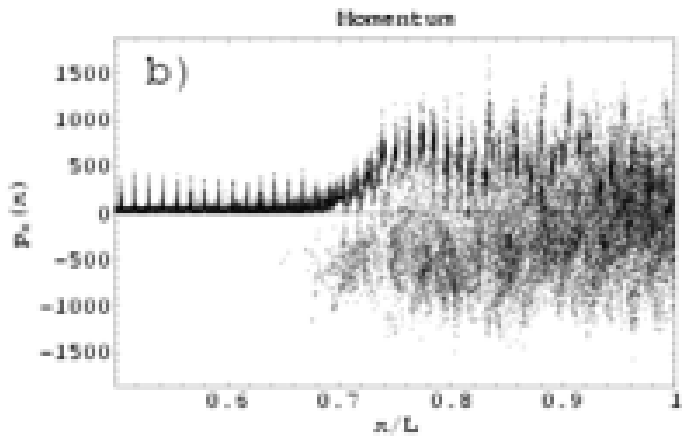} \\
    \includegraphics[scale=0.9]{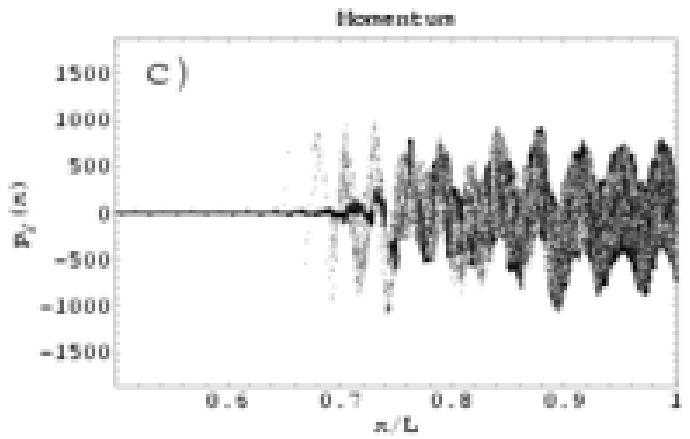} &
    \includegraphics[scale=0.9]{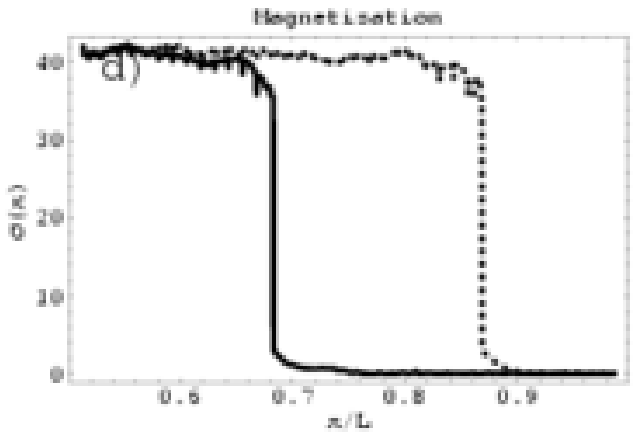} \\
    \includegraphics[scale=0.9]{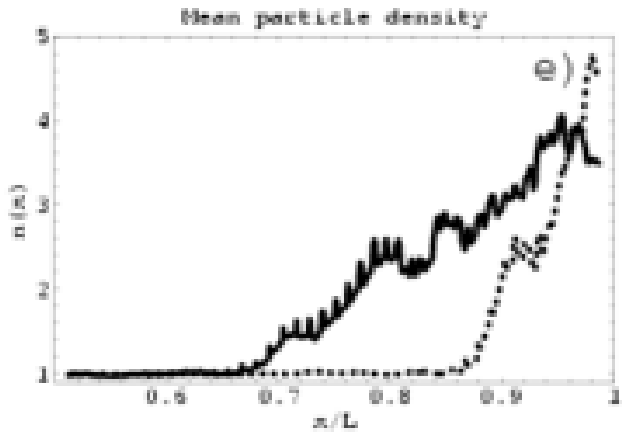}
  \end{tabular}
  \caption{Same as Fig.~\ref{fig:PIC1} but with
    zero average magnetic field, $\alpha=0$.}
  \label{fig:PIC3c}
\end{figure}

Finally, here again, we show an example with zero average magnetic
field $\alpha=0$ in Fig.~\ref{fig:PIC3c}. One sees that the Poynting
flux dissipates completely in this case also. However there is no
sharp shock transition in this case. When the average magnetic field
is nonzero, the shock is mediated by the Larmor rotation of particles
and the width of the shock transition is roughly equal to the Larmor
radius in the average field, $r_0/\alpha$. In our simulations this is
$40/\alpha$ cells so that even at $\alpha=0.1$ the shock transition is
narrow. When $\alpha=0$, the particle free path is determined only by
scattering on magnetic fluctuations therefore the shock transition
could be rather wide.

\subsection{Empirical law}

In the last two paragraphs, we gave two examples of runs, the former
demonstrating that the striped wind can be preserved when some
conditions are fulfilled and the latter showing a case of full
magnetic reconnection. We performed many simulations with different
parameters, changing the magnetisation, the Lorentz factor, the Larmor
radius in order to segregate between the two regimes.

\begin{figure}[htbp]
  \begin{tabular}{cc}
  \includegraphics[scale=1]{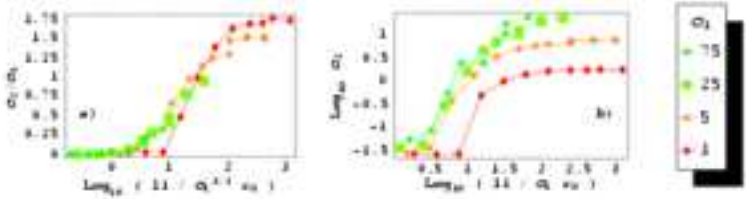}
  \end{tabular}
  \caption{Empirical law for magnetic reconnection in
    the striped wind obtained by 1D PIC simulations. The ratio of the
    downstream to the upstream true magnetisation~$\sigma_2/\sigma_1$
    is plotted against~$l / r_{0} \, \sigma^{3/2}$ in fig~a) and the
    downstream magnetisation against~$l / r_{0} \, \sigma$ in fig~b)
    for different values of the theoretical upstream
    magnetisation~$\sigma$ from mildly~$\sigma=1$ to highly magnetised
    flow~$\sigma=75$.}
  \label{fig:LoiEmpir}
\end{figure}

The results of the full set of simulations can be summarised in plots
showing the ratio of the true downstream to the true upstream
magnetisation~$\sigma_2/\sigma_1$ as well as $\sigma_2$,
Fig.~\ref{fig:LoiEmpir}.  To sum up, we obtain the curves represented
in Fig.~\ref{fig:LoiEmpir}, where these quantities are plotted against
the parameter~$\eta = l / r_{0} \, \sigma^{3/2}$ and $\eta \,
\sqrt{\sigma}$ for fig. a) and b) respectively and for different
initial magnetisations, as was done in the semi-analytical results
presented in Fig.~\ref{fig:Jump1}~e) and~f).

\begin{figure}[htbp]
  \includegraphics[scale=1]{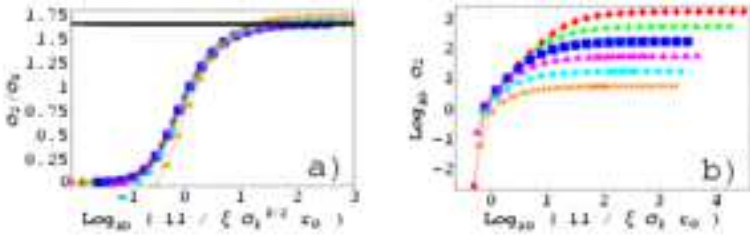}
  \caption{Same figures as Fig.\ref{fig:Jump1}e) and f)
    respectively fig. a) and b) here, but for an adiabatic index
    $\gamma=3/2$ instead of $\gamma=4/3$.}
  \label{fig:Loi}
\end{figure}
For the largest values of~$\eta$, no reconnection is observed as we
would expect from the analytical study discussed in
Sect.~\ref{sec:Solutions}.  In the no dissipation limit, $\eta\gg 1$,
the ratio $\sigma_2/\sigma_1$ reaches values close to 1.7, in
accordance with the expected $5/3$-ratio for the ideal MHD shock with
$\gamma=3/2$ (see appendix~\ref{sec:AnnexeA}). As this limit is
achieved at the same $\eta$ for any $\sigma$, we retrieve the
condition Eq.~(\ref{eq:magom}). According to Fig~\ref{fig:LoiEmpir}~b)
all the presented curves, with the exception of the one corresponding
to $\sigma=1$, go to zero at the same condition
\begin{equation}
  \label{eq:FullDissipPicA}
  \frac{l}{r_0 \, \sigma} \le 3.
\end{equation}
This confirms the analytical criterion \ref{eq:FullDissip} found at
the condition $\sigma\gg 1$.
  
To allow detailed comparisons of the PIC simulations with the
theoretical model presented in sections~\ref{sec:Condition} and
\ref{sec:Solutions}, we plotted in Fig~\ref{fig:Loi} the quantities
$\sigma_2/\sigma_1$ and ~$\sigma_2$ calculated from the theoretical
model with the adiabatic index~ $\gamma=3/2$ (the same quantities were
plotted in Figs.\ref{fig:Jump1}e) and f) for $\gamma=4/3$).
Theoretical jump conditions for arbitrary $\gamma$ are presented in
Appendix B.  Comparing Fig.~\ref{fig:Loi}~a) and
Fig.~\ref{fig:LoiEmpir}~a) on one hand, and Fig.~\ref{fig:Loi}~b) and
Fig.~\ref{fig:LoiEmpir}~b) on the other hand, the parameter~$\xi$ (the
ratio of the sheet width and the particle Larmor radius), introduced
at the end of Sect.~\ref{sec:Condition}, can be easily estimated.
Indeed, looking at fig.\ref{fig:LoiEmpir}a), we find that no
dissipation occurs whenever $l / r_0 \, \sigma^{3/2} \approx 100$
whereas fig.\ref{fig:Loi}a) gives $l / r_0 \, \xi \, \sigma^{3/2}
\approx 10$ so that $\xi \approx 10$. On the other hand,
fig.\ref{fig:LoiEmpir}b) shows full dissipation at $l / r_0 \, \sigma
\approx 3$. Comparing with fig.~\ref{fig:Loi}b) which gives $l / r_0
\, \xi \, \sigma \approx 0.5$, we get $\xi \approx 6$. This means that
$\xi \approx 6-10$.
  
Finally substituting Eqs.(\ref{eq:Magnetisation1}) and
(\ref{eq:Larmoream}) into Eq.(\ref{eq:FullDissipPicA}), the parameter
region for full dissipation is found as
\begin{equation}
  \label{eq:FullDissipPic}
  \mu_0 \, c \, |q| \,
  \frac{l \, n_{\rm c1}}{B_1} \le 3
\end{equation}
Note also that the condition for full dissipation
Eq.~(\ref{eq:FullDissipPic}) is independent of the upstream Lorentz
factor $\Gamma_1$.

\section{Application to pulsar winds}
\label{sec:Wind}

We apply Eq.~(\ref{eq:FullDissipPic}) to pulsar winds. The idea is to
find the limiting radius beyond which the alternating field dissipates
at the termination shock.  Because all quantities refer to the
upstream plasma, we drop the subscript~$1$.

Beyond the light cylinder, we assume that the outflow is {radial}
and propagating in the radial direction. The magnetic field is
predominantly toroidal and decreases with radius as
\begin{equation}
  \label{eq:BL}
  B = B_{\rm L} \, \frac{R_{\rm L}}{R}
\end{equation}
where $R_{\rm L}=c/\Omega$ is the radius of the light cylinder,
$\Omega$ the angular velocity of the neutron star and $B_{\rm L}$ the
magnetic field strength at the light cylinder. The particle number
density falls off as
\begin{equation}
  \label{eq:NL}
  n = n_{\rm L} \, \left( \frac{R_{\rm L}}{R} \right)^2
\end{equation}
where the density at the light cylinder~$n_{\rm L}$ is conventionally
expressed via the so called multiplicity factor~$\kappa$ such that
\begin{equation}
  \label{eq:NL1}
  n_{\rm L} = 2 \, \varepsilon_0 \, \kappa \, \frac{\Omega \, B_{\rm L}}{|q|}
\end{equation}
Half of a wavelength in the striped wind is given by $l = \pi \,
R_{\rm L}$. Substituting these relations into
Eq.(\ref{eq:FullDissipPic}), one finds that full magnetic dissipation
occurs at the termination shock whenever the shock arises at the
distance larger than
\begin{equation}
  \label{eq:DissipCrab}
  \frac{R}{R_{\rm L}} \ge \frac{2\,\pi}{3} \, \kappa. 
\end{equation}
We suppose that the termination shock is stationary in the observer
(pulsar) frame.

The value of the multiplicity, $\kappa$, is rather uncertain.  The
available theoretical models \citep{2001ApJ...554..624H,
  2001ApJ...560..871H}, give $\kappa$ from a few to thousands. The
observed synchrotron emission from the Crab nebula places wide limits
on $\kappa$, from $\kappa\sim 10^4$, at the assumption that only high
energy electrons (those emitting the optical and harder radiation) are
injected now into the nebula, to $\kappa\sim 10^6$ if the radio
emitting electrons are also injected now but not only in the early
stage of the pulsar history. In any case the condition
Eq.~(\ref{eq:DissipCrab}) is easily satisfied for plerions because the
termination shock is located at radii $10^{14}- 10^{16}\,{\rm m}\sim
10^8-10^{10} \, R_{\rm L}$ from the pulsar.  For instance, in the Crab
nebula, the termination shock is located at $R_{\rm shock} / R_{\rm L}
=3 \times 10^9 \gg \kappa$. Thus for these pulsar wind nebulae, the
magnetic energy is easily dissipated at the termination shock so that
the pulsar wind could remain highly magnetised even in the upstream
region close to the shock front.  All the available observation limits
on $\sigma$ are obtained from the analysis of the plasma flow and
radiation beyond the termination shock, the necessary upstream
$\sigma$ being calculated from the ideal MHD jump conditions, see for
instance the recent papers by \cite{2000PASJ...52..875T,
  2001ApJ...561..308S, 2002ApJ...579..404P}.  Our results show that
even if the pulsar wind remains Poynting dominated until the
termination shock, magnetic dissipation at the shock front would
suffice to efficiently convert the electromagnetic energy to the
plasma energy.  This point of view is an alternative to the
$\sigma$-problem.

In binary pulsars, interaction of the pulsar wind with the companion
star results in formation of an intra-binary shock very close to the
neutron star, only thousands of light cylinder radii.
\cite{1993ApJ...403..249A, 1997ApJ...477..439T, 2006ApJ...641..427C}
claim that the X-ray emission observed from the binary pulsars
PSR~1957+20 and PSR~1259-63 is explained if the magnetisation
downstream of the shock is small so that the Poynting flux is already
converted into the plasma energy.  Then we could estimate from
Eq.(\ref{eq:DissipCrab}) the upper limit of the pair production
factor~$\kappa$ for these pulsars.

Indeed, first consider the pulsar PSR~1957+20. It has a rotation
period of $P=1.607$~ms, corresponding to a light cylinder radius of
$R_{\rm L} = 76.7$~km.  The intra-binary shock arises at a distance
less than the orbital separation to the companion star, so that we get
$R_{\rm shock} \lesssim 1.7 \times 10^{6}$~km.  According to
Eq.~(\ref{eq:DissipCrab}), in order to expect a significant
dissipation of the magnetic field at the shock, the multiplicity
factor should then satisfy $\kappa \le ( 1 / \pi ) \, R_{\rm
  shock} / R_{\rm L} \approx \times 10^4$.

Next consider the pulsar PSR~1259-63. It possesses a period
of~$P=47.7$~ms, corresponding to a light cylinder radius $R_{\rm
  L}=2279$~km. Because here again the shock should arise at a distance
smaller than the orbital separation, we found $R_{\rm shock} \lesssim
3.9\times 10^{8}$~km. Magnetic reconnection implies $\kappa \lesssim
8 \times 10^4$.

Another interesting case is the famous double pulsar PSR J0737-3039.
According to \cite{2004ApJ...613L..57M}, the radio emission from the
pulsar~B is modulated with the period of pulsar~A.  This strongly
supports the idea that the magnetosphere of pulsar~B is disturbed by
the striped wind emanating from pulsar~A.  The condition that the
striped structure has not been erased at the distance separating both
pulsars, imposes a lower limit on the multiplicity factor~$\kappa$ in
this case.  Let us estimate this limit. The period of pulsar~A is
$P_A=22.699$~ms corresponding to a light cylinder radius of~$R_{\rm L}
= 1083$~km.  The separation between the two pulsars is approximately
$700,000$~km \citep{2005PhysicsWorld...29L}.  Then the lower limit for
the ratio in Eq.~(\ref{eq:DissipCrab}) is $R/R_{\rm L} \approx 646$.
We conclude that the lower limit for the pair production multiplicity
in pulsar~A is~$\kappa \gtrsim 310$.

\section{CONCLUSION}
\label{sec:Conclusion}

In this paper, we studied the magnetic reconnection process at the
termination shock in the pulsar striped wind. Using the jump
conditions for the averaged quantities (over a wavelength) in the
striped wind, we derived a simple analytical criterion for magnetic
dissipation, see Table~\ref{tab:Critere}.  However, we had to
introduce a free parameter~$\xi$, Eq.~(\ref{eq:Delta2}), entering the
dissipation condition. It relates the downstream thickness~$\Delta_2$
of a current sheet to the downstream Larmor radius~$r_{0}$ by the
following expression~$\Delta_2 = \xi \, r_{0}$. $\xi$ is the only free
parameter in our model.  In order to estimate this parameter, we
performed 1D relativistic PIC simulations and found that $\xi \approx
10$.

Knowing the important flow parameters in the incoming plasma which
are half the wavelength~$l$, the magnetisation~$\sigma$ and the
Larmor radius~$r_{0}$, we are able to predict the percentage of
magnetic reconnection in the striped wind when crossing the
termination shock, according to the following law
\begin{itemize}
\item for $ l / r_{0} \, \sigma \le 3 $, full dissipation occurs,
  the flow downstream is purely hydrodynamical with Lorentz factor
  close to unity~$\Gamma_2\approx1$, particle are thermalised and
  heated to relativistic temperature. The magnetic field has
  completely disappeared, except for the DC component (i.e. average
  magnetic field)~;
\item for $\sigma \le ( l / 12 \, r_{0} )^{2/3}$, no magnetic
  reconnection exists. The striped wind structure is preserved
  downstream, it is just compressed. The downstream Lorentz factor is
  the same as for ideal MHD, $\Gamma_2 = \sqrt{\sigma}$~;
\item $ 3 \le l / r_{0} \, \sigma \le 12 \, \sqrt{\sigma} $, the
  magnetic field is partially dissipated. The stripes are weakened and
  the flow is decelerated to a downstream Lorentz factor~$\Gamma_2 = l
  / (12 \, r_{0} \, \sigma)$.
\end{itemize}

We applied the condition for full magnetic dissipation,
Eq.~(\ref{eq:DissipCrab}), to pulsar wind nebula and binary
pulsars. Because in plerions the termination shock is located at
radii $10^{14}- 10^{16}\,{\rm m}\sim 10^8-10^{10} \, R_{\rm L}$
from the pulsar, magnetic reconnection is easily achieved. This
conclusion could resolve a long-standing difficulty with
transformation of the electro-magnetic energy of the pulsar wind
into the plasma energy in the nebula (the so called
$\sigma$-problem). First of all, both observations of the X-ray
tori \citep{2000ApJ...536L..81W} and theoretical models
\citep{1999A&A...349.1017B} suggest that most of the energy in the
pulsar wind is transferred in the equatorial belt where the
magnetic field is predominantly alternating. We see now that even
though dissipation of the alternating fields in the wind is
hampered by relativistic slowing-down of time
\citep{2001ApJ...547..437L, 2003ApJ...591..366K}, the
electro-magnetic energy is readily released at the terminating
shock.

Our model could be applied also to pulsars in binary systems, where
interaction of the pulsar wind with the companion star results in a
formation of a shock relatively close to the pulsar.  In this case,
the presence or absence of magnetic dissipation imposes a strong
constraint on the pair multiplicity factor~$\kappa$. For pulsars
PSR~1259-63 and PSR~1957+20, where full dissipation is expected, the
upper limit is roughly $\kappa \lesssim {\rm few} \times 10^4$. For
the binary pulsar PSR~0737-3039, we do not expect full reconnection
and therefore the lower limit is found to be $\kappa \gtrsim 310$.

The generation of electromagnetic waves at the shock front and their
propagation in the upstream plasma will affect the flow already before
entering the discontinuity. This aspect of the reconnection in the
termination shock is left for future work.

\begin{acknowledgements}
  We are grateful to the anonymous referee for his insightful comments
  and valuable suggestions. This work was supported by a grant from
  the G.I.F., the German-Israeli Foundation for Scientific Research
  and Development.
\end{acknowledgements}


\begin{thebibliography}{43}
\expandafter\ifx\csname natexlab\endcsname\relax\def\natexlab#1{#1}\fi

\bibitem[{{Arons} \& {Tavani}(1993)}]{1993ApJ...403..249A}
{Arons}, J. \& {Tavani}, M. 1993, \apj, 403, 249

\bibitem[{{Beskin} {et~al.}(1998){Beskin}, {Kuznetsova}, \&
  {Rafikov}}]{1998MNRAS.299..341B}
{Beskin}, V.~S., {Kuznetsova}, I.~V., \& {Rafikov}, R.~R. 1998, \mnras, 299,
  341

\bibitem[{{Birdsall} \& {Langdon}(2005)}]{Birdsall2005}
{Birdsall}, C. \& {Langdon}, A. 2005, {Plasma physics via computer simulation}
  (IOP Publishing)

\bibitem[{{Bogovalov} \& {Tsinganos}(1999)}]{1999MNRAS.305..211B}
{Bogovalov}, S. \& {Tsinganos}, K. 1999, \mnras, 305, 211

\bibitem[{{Bogovalov}(1999)}]{1999A&A...349.1017B}
{Bogovalov}, S.~V. 1999, \aap, 349, 1017

\bibitem[{{Cheng} {et~al.}(2006){Cheng}, {Taam}, \&
  {Wang}}]{2006ApJ...641..427C}
{Cheng}, K.~S., {Taam}, R.~E., \& {Wang}, W. 2006, \apj, 641, 427

\bibitem[{{Chiueh} {et~al.}(1998){Chiueh}, {Li}, \&
  {Begelman}}]{1998ApJ...505..835C}
{Chiueh}, T., {Li}, Z.-Y., \& {Begelman}, M.~C. 1998, \apj, 505, 835

\bibitem[{{Coroniti}(1990)}]{1990ApJ...349..538C}
{Coroniti}, F.~V. 1990, \apj, 349, 538

\bibitem[{{Emmering} \& {Chevalier}(1987)}]{1987ApJ...321..334E}
{Emmering}, R.~T. \& {Chevalier}, R.~A. 1987, \apj, 321, 334

\bibitem[{{Gallant} {et~al.}(1992){Gallant}, {Hoshino}, {Langdon}, {Arons}, \&
  {Max}}]{1992ApJ...391...73G}
{Gallant}, Y.~A., {Hoshino}, M., {Langdon}, A.~B., {Arons}, J., \& {Max}, C.~E.
  1992, \apj, 391, 73

\bibitem[{{Hibschman} \& {Arons}(2001{\natexlab{a}})}]{2001ApJ...554..624H}
{Hibschman}, J.~A. \& {Arons}, J. 2001{\natexlab{a}}, \apj, 554, 624

\bibitem[{{Hibschman} \& {Arons}(2001{\natexlab{b}})}]{2001ApJ...560..871H}
{Hibschman}, J.~A. \& {Arons}, J. 2001{\natexlab{b}}, \apj, 560, 871

\bibitem[{{Hoshino} {et~al.}(1992){Hoshino}, {Arons}, {Gallant}, \&
  {Langdon}}]{1992ApJ...390..454H}
{Hoshino}, M., {Arons}, J., {Gallant}, Y.~A., \& {Langdon}, A.~B. 1992, \apj,
  390, 454

\bibitem[{{Kennel} \& {Coroniti}(1984{\natexlab{a}})}]{1984ApJ...283..694K}
{Kennel}, C.~F. \& {Coroniti}, F.~V. 1984{\natexlab{a}}, \apj, 283, 694

\bibitem[{{Kennel} \& {Coroniti}(1984{\natexlab{b}})}]{1984ApJ...283..710K}
{Kennel}, C.~F. \& {Coroniti}, F.~V. 1984{\natexlab{b}}, \apj, 283, 710

\bibitem[{{Kirk}(2005)}]{2005PPCF...47B.719K}
{Kirk}, J.~G. 2005, Plasma Physics and Controlled Fusion, 47, B719

\bibitem[{{Kirk} \& {Skj{\ae}raasen}(2003)}]{2003ApJ...591..366K}
{Kirk}, J.~G. \& {Skj{\ae}raasen}, O. 2003, \apj, 591, 366

\bibitem[{{Levinson} \& {van Putten}(1997)}]{1997ApJ...488...69L}
{Levinson}, A. \& {van Putten}, M.~H.~P.~M. 1997, \apj, 488, 69

\bibitem[{{Lyne} \& {Kramer}(2005)}]{2005PhysicsWorld...29L}
{Lyne}, A.~G. \& {Kramer}, M. 2005, Physics World, 29

\bibitem[{{Lyubarsky}(2005{\natexlab{a}})}]{2005xrrc.procE5.02L}
{Lyubarsky}, Y. 2005{\natexlab{a}}, in X-Ray and Radio Connections (eds. L.O.
  Sjouwerman and K.K Dyer) Published electronically by NRAO,
  http://www.aoc.nrao.edu/events/xraydio Held 3-6 February 2004 in Santa Fe,
  New Mexico, USA, (E5.02) 10 pages

\bibitem[{{Lyubarsky}(2005{\natexlab{b}})}]{2005AdSpR..35.1112L}
{Lyubarsky}, Y. 2005{\natexlab{b}}, Advances in Space Research, 35, 1112

\bibitem[{{Lyubarsky} \& {Eichler}(2001)}]{2001ApJ...562..494L}
{Lyubarsky}, Y. \& {Eichler}, D. 2001, \apj, 562, 494

\bibitem[{{Lyubarsky} \& {Kirk}(2001)}]{2001ApJ...547..437L}
{Lyubarsky}, Y. \& {Kirk}, J.~G. 2001, \apj, 547, 437

\bibitem[{{Lyubarsky}(2003)}]{2003MNRAS.345..153L}
{Lyubarsky}, Y.~E. 2003, \mnras, 345, 153

\bibitem[{{McLaughlin} {et~al.}(2004){McLaughlin}, {Kramer}, {Lyne}, {Lorimer},
  {Stairs}, {Possenti}, {Manchester}, {Freire}, {Joshi}, {Burgay}, {Camilo}, \&
  {D'Amico}}]{2004ApJ...613L..57M}
{McLaughlin}, M.~A., {Kramer}, M., {Lyne}, A.~G., {et~al.} 2004, \apjl, 613,
  L57

\bibitem[{{Michel}(1971)}]{1971CoASP...3...80M}
{Michel}, F.~C. 1971, Comments on Astrophysics and Space Physics, 3, 80

\bibitem[{{Michel}(1973)}]{1973ApJ...180L.133M}
{Michel}, F.~C. 1973, \apjl, 180, L133+

\bibitem[{{Michel}(1982)}]{1982RvMP...54....1M}
{Michel}, F.~C. 1982, Reviews of Modern Physics, 54, 1

\bibitem[{{Michel}(1991)}]{1991tnsm.book.....M}
{Michel}, F.~C. 1991, {Theory of neutron star magnetospheres} (Chicago, IL,
  University of Chicago Press, 1991, 533 p.)

\bibitem[{{Michel}(1994)}]{1994ApJ...431..397M}
{Michel}, F.~C. 1994, \apj, 431, 397

\bibitem[{{Michel}(2005)}]{2005RMxAC..23...27M}
{Michel}, F.~C. 2005, in Revista Mexicana de Astronomia y Astrofisica
  Conference Series, 27--34

\bibitem[{{Petre} {et~al.}(2002){Petre}, {Kuntz}, \&
  {Shelton}}]{2002ApJ...579..404P}
{Petre}, R., {Kuntz}, K.~D., \& {Shelton}, R.~L. 2002, \apj, 579, 404

\bibitem[{{P{\'e}tri} \& {Kirk}(2005)}]{2005ApJ...627L..37P}
{P{\'e}tri}, J. \& {Kirk}, J.~G. 2005, \apjl, 627, L37

\bibitem[{{Piran}(2005)}]{2005RvMP...76.1143P}
{Piran}, T. 2005, Reviews of Modern Physics, 76, 1143

\bibitem[{{Rees} \& {Gunn}(1974)}]{1974MNRAS.167....1R}
{Rees}, M.~J. \& {Gunn}, J.~E. 1974, \mnras, 167, 1

\bibitem[{{Safi-Harb} {et~al.}(2001){Safi-Harb}, {Harrus}, {Petre}, {Pavlov},
  {Koptsevich}, \& {Sanwal}}]{2001ApJ...561..308S}
{Safi-Harb}, S., {Harrus}, I.~M., {Petre}, R., {et~al.} 2001, \apj, 561, 308

\bibitem[{{Spitkovsky}(2006)}]{2006ApJ...648L..51S}
{Spitkovsky}, A. 2006, \apjl, 648, L51

\bibitem[{{Tavani} \& {Arons}(1997)}]{1997ApJ...477..439T}
{Tavani}, M. \& {Arons}, J. 1997, \apj, 477, 439

\bibitem[{{Tomimatsu}(1994)}]{1994PASJ...46..123T}
{Tomimatsu}, A. 1994, \pasj, 46, 123

\bibitem[{{Torii} {et~al.}(2000){Torii}, {Slane}, {Kinugasa}, {Hashimotodani},
  \& {Tsunemi}}]{2000PASJ...52..875T}
{Torii}, K., {Slane}, P.~O., {Kinugasa}, K., {Hashimotodani}, K., \& {Tsunemi},
  H. 2000, \pasj, 52, 875

\bibitem[{{Usov}(1975)}]{1975Ap&SS..32..375U}
{Usov}, V.~V. 1975, \apss, 32, 375

\bibitem[{{Weisskopf} {et~al.}(2000){Weisskopf}, {Hester}, {Tennant}, {Elsner},
  {Schulz}, {Marshall}, {Karovska}, {Nichols}, {Swartz}, {Kolodziejczak}, \&
  {O'Dell}}]{2000ApJ...536L..81W}
{Weisskopf}, M.~C., {Hester}, J.~J., {Tennant}, A.~F., {et~al.} 2000, \apjl,
  536, L81

\bibitem[{{Zhang} \& {Kobayashi}(2005)}]{2005ApJ...628..315Z}
{Zhang}, B. \& {Kobayashi}, S. 2005, \apj, 628, 315

\end{thebibliography}

\appendix

\section{Jump conditions for arbitrary adiabatic index~$\gamma$}
\label{sec:AnnexeA}

\subsection{Ideal MHD}

In this appendix, we give the general expressions for the ideal MHD
jump conditions for arbitrary adiabatic index~$\gamma$,
\citep{1992ApJ...391...73G, 1992ApJ...390..454H}.  Let's assume that
the equation of state for the ultrarelativistic plasma is given by
\begin{equation}
  \label{eq:EOS}
  e = \frac{p}{\gamma - 1}
\end{equation}
for an arbitrary index~$\gamma$. The enthalpy is therefore given by
\begin{equation}
  \label{eq:Enthalpy}
  w = \frac{\gamma \, p}{\gamma-1}
\end{equation}
Following the procedure of \cite{1984ApJ...283..694K}, the solutions
to the MHD jump conditions for the 4-velocity gives
\begin{eqnarray}
  \label{eq:Quadvit}
  u_2^2 & = & \frac{ ( 4 - \gamma )\, \gamma \, \sigma_1^2 +
    4 \, ( 1 + (\gamma-1)^2 ) \, \sigma_1 + 4 \, (\gamma-1)^2}
  {8 \, \gamma ( 2 - \gamma ) \, ( \sigma_1 + 1 )} \\
  & & \pm \frac{\sqrt{ 16 \, \gamma \, (\gamma-2)^3 \, \sigma_1^2 \, (\sigma_1+1) +
      [ (\gamma-2)^2 \, \sigma_1^2 - 4 \, (\sigma_1+1) \, ( \sigma_1 + (\gamma-1)^2 )]^2} }
  {8 \, \gamma ( 2 - \gamma ) \, ( \sigma_1 + 1 )} \nonumber
\end{eqnarray}
Starting from the 4-velocity, we can compute the Lorentz factor, the
downstream temperature and the magnetisation by
\begin{eqnarray}
  \Gamma_2 & = & \sqrt{ 1 + u_2^2 } \\
  \frac{k_B \, T_2}{\Gamma_1 \, m \, c^2} & = &
  \frac{\gamma-1}{\gamma} \, \frac{1}{\Gamma_2} \,
  \left[ 1 + \sigma_1 \, \left( 1 - \frac{1}{\beta_2 } \right) \right] \\
  \sigma_2 & = & \frac{\gamma-1}{\gamma} \, \frac{n_2}{n_1} \,
  \frac{\Gamma_1}{\Gamma_2} \, \frac{m\,c^2}{k_B \, T_2} \, \sigma_1 \\
  \frac{n_2}{n_1} & = & \frac{1}{\beta_2}
\end{eqnarray}
For $\gamma=4/3$ we retrieve the usual
result
\begin{equation}
  \label{eq:u2KC}
  u_2^2 = \frac{8 \, \sigma_1^2 + 10 \, \sigma_1 + 1 \pm
    \sqrt{64 \, \sigma_1^4 + 128 \, \sigma_1^3 + 84 \, \sigma_1^2 + 20 \, \sigma_1 +1}}
  {16 \, (\sigma_1 + 1 )}
\end{equation}
For arbitrary $\gamma$, the expansion in $1/\sigma_1$ to the second order
leads to
\begin{equation}
  \label{eq:u2c2e}
  u_2^2 = \frac{\gamma-4}{4\,(\gamma-2)} \, \sigma_1 -
  \frac{\gamma^2 - 8 \, \gamma + 8}{4 \, \left(\gamma^2 - 6 \, \gamma + 8 \, \right)}
\end{equation}
The corresponding Lorentz factor is to the same order
\begin{equation}
  \label{eq:G2gam}
  \Gamma_2 = \frac{1}{2} \,
  \sqrt{\frac{(\sigma_1 + 3) \, \gamma^2 - 8 \, (\sigma_1 +2) \, \gamma +
      8 \, (2 \sigma_1 + 3)}{\gamma^2 - 6 \, \gamma +8}}
\end{equation}
This leads to a downstream temperature
\begin{equation}
  \label{eq:T2gam}
  \frac{k_B \, T_2}{\Gamma_1 \, m \, c^2} =
  \frac{\gamma-1}{4-\gamma} \, \sqrt{\frac{\gamma-2}{\gamma-4}} \,
  \frac{1}{\sqrt{\sigma_1}} \, \left( 2 -
    \frac{\gamma^3 \,
      \left(3 \, \gamma^2 - 8 \, \gamma + 8 \right)}
    {(\gamma-4)^2 \, \gamma^3} \, \frac{1}{\sigma_1} \right)
\end{equation}
a density jump
\begin{equation}
  \label{eq:n2sn1}
  \frac{n_2}{n_1} = 1 + 2\, \frac{\gamma - 2}{\gamma - 4} \, \frac{1}{\sigma_1}
\end{equation}
and a magnetisation
\begin{equation}
  \label{eq:s2ss1}
  \sigma_2 = \left( \frac{4-\gamma}{\gamma} \, -
    2 \, \frac{(\gamma-2)^2}{\gamma \, (\gamma-4) \, \sigma_1} \right) \, \sigma_1
\end{equation}

In Table~\ref{tab:Index}, we give the numerical values for the
different coefficients for $\gamma=4/3$ and $\gamma=3/2$.

\begin{table}[htbp]
  \centering
  \begin{tabular}{ccc}
    \hline
    Quantity & $\gamma=4/3$ & $\gamma=3/2$ \\
    \hline
    \hline
    $\Gamma_2$ & $\sqrt{\sigma_1+\frac{9}{8}}$ &
    $\sqrt{\frac{5}{4} \, \sigma_1 + \frac{27}{20}}$ \\
    $k_B \, T_2/m \, c^2$ & $\frac{\Gamma_1}{16\,\sqrt{\sigma_1}} \,
    \left( 2 - \frac{3}{8\,\sigma_1} \right)$ &
    $\frac{\Gamma_1}{5\,\sqrt{5\,\sigma_1}} \,
    \left( 2 - \frac{11}{25\,\sigma_1} \right)$\\
    $n_2/n_1$ & $ 1 + \frac{1}{2\,\sigma_1} $ & $ 1 + \frac{2}{5\,\sigma_1} $ \\
    $\sigma_2/\sigma_1$ & $ 2 + \frac{1}{4\,\sigma_1} $ &
    $ \frac{5}{3} + \frac{2}{15\,\sigma_1} $ \\
    \hline
  \end{tabular}
  \caption{Downstream parameters for the
    two adiabatic indexes~$\gamma=4/3,3/2$,
    expanded to first order in the upstream magnetisation~$1/\sigma_1$.}
  \label{tab:Index}
\end{table}

\subsection{Striped wind}

We now give the general formulae for the jump conditions in the
striped wind for arbitrary adiabatic index~$\gamma$.  Following the
same procedure as in Sect~\ref{sec:Condition} using the general
expression described in the previous section, the average conservation
of particle and energy-momentum read,
\begin{eqnarray}
  \label{eq:CsvPartGen}
  \beta_1 \, \left[ ( 1 - \delta_1 ) \, n_{\rm c1} + \delta_1 \, n_{\rm h1} \right] & = &
  \beta_2 \, \left[ ( 1 - \delta_2 ) \, n_{\rm c2} + \delta_2 \, n_{\rm h2} \right] \\
  \label{eq:CsvEnerGen}
  \beta_1 \, n_{\rm c1} \, \left[ 1 - \delta_1 +
    \left( \frac{2-\gamma}{2\,(\gamma-1)} \, \delta_1 + 1 \right) \, \sigma_1 \right] & = &
  \beta_2 \, n_{\rm c2} \, \left[ \frac{\gamma}{4-\gamma} +
    \left( 1 + \frac{2-\gamma}{2\,(\gamma-1)} \, \delta_2 \right) \,
    \frac{n_{\rm c2}}{n_{\rm c1}} \, \sigma_1 \right] \\
  \label{eq:CsvImpGen}
  n_{\rm c1} \, \left[ \Gamma_1 \, \beta_1^2 \, \left( 1 - \delta_1 +
      \left( \frac{2-\gamma}{2\,(\gamma-1)} \, \delta_1 + 1 \right) \, \sigma_1 \right)
    + \frac{\sigma_1}{2\,\Gamma_1} \right] & = &
  \frac{\Gamma_1}{\Gamma_2} \, n_{\rm c2} \, \left[ \Gamma_2 \, \beta_2^2
    \, \left( \frac{\gamma}{4-\gamma} +
      \left( 1 + \frac{2-\gamma}{2\,(\gamma-1)} \, \delta_2 \right) \,
      \frac{n_{\rm c2}}{n_{\rm c1}} \, \sigma_1 \right) + \right. \nonumber \\
  & & \left. \frac{\gamma-1}{4-\gamma}
    \, \frac{1}{\Gamma_2} + \frac{n_{\rm c2} \, \sigma_1}{2\,n_{\rm c1}\,\Gamma_2} \right]
\end{eqnarray}
The prescription for the current sheet thickness downstream is
\begin{equation}
  \label{eq:D2Gen}
  \delta_2 = \xi \, \left[ \frac{\gamma-1}{4-\gamma} +
    \frac{n_{\rm c2}\,\sigma_1}{2\,n_{\rm c1}} \right] \,
  \frac{\beta_1 \, n_{\rm c1}}{\Gamma_2 \, \beta_2 \, n_{\rm h2}} \, \frac{r_0}{l_1}
\end{equation}

\end{document}